\documentclass[10pt, journal]{IEEEtran}

\usepackage[utf8x]{inputenc} 
\usepackage{amsmath}
\usepackage{amsfonts}
\usepackage{amssymb}
\usepackage{paralist}
\usepackage{graphicx}
\usepackage{booktabs}
\usepackage{bm}
\usepackage{algpseudocode}
\usepackage{multicol}
\usepackage{stfloats}
\usepackage{url}
\usepackage{enumitem}

\usepackage[table]{xcolor}
\usepackage{threeparttable}
\usepackage{multirow}

\usepackage[caption=false,font=footnotesize]{subfig}
\usepackage[level]{datetime}  
\newdateformat{ukdate}{\ordinaldate{\THEDAY} \monthname[\THEMONTH] \THEYEAR}

\usepackage{hyperref}
\usepackage[nameinlink,noabbrev]{cleveref}
\usepackage[square, comma, sort&compress, numbers]{natbib}

\usepackage[ruled,vlined]{algorithm2e}
\usepackage{setspace}
\SetAlgoSkip{}
\SetAlgoInsideSkip{}
\SetAlFnt{\small}
\SetAlCapFnt{\small}
\SetAlCapNameFnt{\small}
\SetKwRepeat{Do}{do}{while}
\SetEndCharOfAlgoLine{}
\SetAlgoCaptionSeparator{}

\usepackage{color}

\graphicspath{{figures/}}

\IEEEaftertitletext{\vspace{-1.5\baselineskip}}

\usepackage{amsthm}

\newtheorem{defn}{Definition}
\newtheorem{rem}{Remark} 
\makeatletter

\begin{document}

\title{Channel Estimation and Hybrid Precoding for Distributed Phased Arrays Based MIMO \\Wireless Communications}
%
%
%

\author{Yu Zhang,~\IEEEmembership{Student Member,~IEEE,}  Yiming Huo,~\IEEEmembership{Member,~IEEE,} 
Dongming Wang,~\IEEEmembership{Member,~IEEE,}  \\
Xiaodai Dong,~\IEEEmembership{Senior~Member,~IEEE,}  
and Xiaohu You,~\IEEEmembership{Fellow,~IEEE}
\thanks{This work was supported in part by the National Key Research and Development Program under Grant 2018YFE0205902, the National Natural Science Foundation of China (NSFC) under Grant 61871122 and Grant 61971127, and the Six Talent Peaks Project in Jiangsu Province. (\emph{Corresponding authors: Xiaodai Dong; Dongming Wang)}}
\thanks{Y. Zhang, D. Wang and X. You are with the National Mobile Communications Research Laboratory, Southeast University, Nanjing 210096, China (email: \{yuzhang,wangdm,xhyu\}@seu.edu.cn). D. Wang and X. You are also with the Purple Mountain Laboratories, Nanjing 211111, China.}
\thanks{Y. Huo and X. Dong are with the Department of Electrical and Computer Engineering,  University of Victoria, Victoria, BC V8P 5C2, Canada, email: (ymhuo@uvic.ca, xdong@ece.uvic.ca).}
}

%
%

\markboth{IEEE Transactions on Vehicular Technology}%
{}
%




\maketitle

\begin{abstract}
Distributed phased arrays based multiple-input multiple-output (DPA-MIMO) is a newly introduced architecture that enables both spatial multiplexing and beamforming while facilitating highly reconfigurable hardware implementation in millimeter-wave (mmWave) frequency bands. With a DPA-MIMO system, we focus on channel state information (CSI) acquisition and hybrid precoding. As benefited from a coordinated and open-loop pilot beam pattern design, all the sub-arrays  can  perform channel sounding with less training overhead compared with the traditional orthogonal operation of each sub-array. Furthermore, two sparse channel recovery algorithms, known as joint orthogonal matching pursuit (JOMP) and joint sparse Bayesian learning with $\ell_2$ reweighting (JSBL-$\ell_2$), are proposed to exploit the hidden structured sparsity in the beam-domain channel vector. Finally, successive interference cancellation (SIC) based hybrid precoding through sub-array grouping is illustrated for the DPA-MIMO system, which decomposes the joint sub-array RF beamformer design into an interactive per-sub-array-group handle. Simulation results show that the proposed two channel estimators fully take advantage of the partial coupling characteristic of DPA-MIMO channels to perform channel recovery, and the proposed hybrid precoding algorithm is suitable for such array-of-sub-arrays architecture with satisfactory performance and low complexity. 
\end{abstract}

\begin{IEEEkeywords}
Distributed phased arrays based multiple-input multiple-output (DPA-MIMO), millimeter-wave (mmWave), array-of-sub-arrays, channel estimation, orthogonal matching pursuit (OMP), sparse Bayesian learning (SBL), successive interference cancellation (SIC), hybrid precoding.
\end{IEEEkeywords}

%
\IEEEpeerreviewmaketitle

\section{Introduction}
\IEEEPARstart{D}{riven} by the tremendous growth in demand for wireless data, many new technologies have been proposed  for fifth generation cellular communications (5G) to enable orders of magnitude increases in the network capacity~\cite{rappaport2013millimeter,li2018mmwave}. In the physical layer, the exploration of new spectrum in the so-called 5G upper bands, for example, from 6 GHz up to 100 GHz, including the millimeter-wave (mmWave) frequencies, has made multi-gigabit-per-second wireless communications more promising and feasible~\cite{raghavan2018millimeter}. However, applying mmWave communications to commercial cellular networks is very challenging mainly due to, first, much higher propagation losses compared with those at lower microwave frequencies; second, strict constraints on hardware designs and implementations which include but are not limited to, antenna performance and dimension, circuits and systems integration challenges~\cite{huo2018cellular}, power consumption and power supply, form factor (particularly critical for a mobile handset device), etc., according to~\cite{huo20175g}. Fortunately,  massive antenna elements working at mmWave bands can be accommodated into a limited hardware area due to shorter wavelength, facilitating large beamforming gain for combating large pathloss and establishing stable links with reasonable signal-to-noise ratio (SNR) values.

To enable massive multiple-input multiple-output (MIMO) communication with less  ratio frequency (RF) chains, a hybrid analog-digital solution~\cite{el2014spatially,liang2014low} was proposed. In this architecture,  the signal processing in conventional MIMO is divided into low-dimensional digital beamforming, and high-dimensional analog beamforming that is implemented with low-cost phase shifters.
This hybrid transceiver topology is further categorized into the fully-connected and sub-array based structures in terms of how RF chains are mapped to antennas. In a fully-connected structure, each RF chain enables full array gain through individual connection to all antennas~\cite{raghavan2017single}; while for the latter structure, each RF chain is only connected to partial antennas, which reduces complexity at the penalty of degrading beamforming gain~\cite{gao2016energy}. 
In practice, the array-of-sub-array structure has drawn great attention due to its low-complexity implementation, high energy efficiency, and flexible configurations~\cite{huang2010hybrid,singh2015feasibility,park2017dynamic}.

Recently, a distributed phased arrays based MIMO (DPA-MIMO) architecture which can be easily applied to both base station (BS) and user equipment (UE) designs, has been proposed for practical system and hardware designs~\cite{huo20175g}. One of its key characteristics distinguishing from the traditional array-of-sub-array structure is to deploy sub-arrays in separate locations similar to distributed antenna systems~\cite{wang2013spectral} with relatively small separation.
More importantly, the underlying mechanism of DPA-MIMO requires the sub-arrays to separate from each other for several reasons~\cite{huo2018cellular,huo20175g,huo2019enabling}: 1) The practical coupling effects that cause spatial interference (such as the unavoidable side lobes) can be mitigated to guarantee the isolation and independence of each sub-array; 2) The heat dissipation capability and thermal performance can be enhanced by separating sub-arrays with some distance; 3) In particular, at the UE end, separating sub-arrays is critical to overcome the human body (hand) blockage issue. This highly reconfigurable architecture facilitates the multi-beam multi-stream based 5G system and hardware designs under some realistic constraints and resources limitation, which enables appealing advanced features for both academic research and industrial applications~\cite{cuvelier2018mmwave,huo2019multi,huo2019distributed,zhang2019admm}. 

Owing to the geographically separated sub-array structure, the DPA-MIMO based channel modeling requires an in-depth exploration. The basic idea has been inspired by the channel measurements for outdoor BS composed of  a very large array~\cite{payami2012channel}.
Instead, this extra-large MIMO channel cannot be characterized as a wide sense stationary (WSS) model in the spatial domain. Following the analysis in~\cite{payami2012channel,gao2015massive}, this spatially non-WSS channel feature results from that, some clusters are only visible to a part of the large array due to some practical factors, such as the cluster sizes, the array aperture, and the spacing between these clusters and the array. 
Furthermore, this spatial non-stationarity in massive MIMO channels was visualized by introducing the concepts of partially visible clusters and wholly visible clusters which are categorized according to their visibility regions (VRs)~\cite{li2015capacity,ali2019linear}.
A recent measurement campaign has further verified this spatially non-stationary characteristic over a $40\times 40$ planar RX array at frequencies from 13 to 17 GHz~\cite{chen2017measurement}. 
Based on the above discussion, increasing the spacing between adjacent sub-arrays 
inevitably triggers an effect that independent scatterers appear over different sub-arrays in a DPA-MIMO system due to their  VRs, which has also occurred in indoor THz communications~\cite{lin2015adaptive}. 
Meanwhile, as constrained by hardware dimension and power consumption, a reasonable distance between adjacent sub-arrays should be set~\cite{huo20175g}. As a result, the channels between any transmitter (TX) and receiver (RX) sub-arrays are partial coupling, i.e.,  both common and local scatterers exist in a DPA-MIMO transmission environment. Due to the inherent sparsity of mmWave channels, we consider modeling the DPA-MIMO channel using the virtual angular domain representation~\cite{sayeed2002deconstructing}. In this way, the channel coupling relationship can be conveniently characterized as structured sparsity in the beam domain~\cite{gao2018compressive}.

Channel state information (CSI) acquisition in mmWave systems is challenging due to high dimensional channels, and low point-to-point SNR before beamforming. This makes conventional channel estimators such as the least square (LS) approach infeasible. Exploiting the mmWave channel sparsity to reduce the training overhead is expected to address this challenge~\cite{sun2017millimeter,gao2018compressive,huang2018iterative}. By appropriately choosing the RF and digital precoder matrices, the work in~\cite{lee2016channel} developed a spatial grid based orthogonal matching pursuit (OMP) method to estimate the channel of hybrid MIMO systems. A sparse Bayesian learning (SBL) based channel estimator in~\cite{mishra2017sparse} was further demonstrated to achieve better performance than the greedy method that is sensitive to the choice of the dictionary matrix in~\cite{lee2016channel}.  By leveraging the common sparsity over multiple measurement vectors (MMV), the authors in~\cite{srivastava2019quasi} applied an SBL based approach to mmWave hybrid MIMO systems for accurate support detection.  This type of block sparsity, e.g., the common angle-domain channel sparsity over all pilot subcarriers, has also been observed and exploited in~\cite{gao2018compressive,ke2020compressive} to enhance broadband channel estimation when orthogonal frequency-division multiplexing (OFDM) modulation is utilized. However, different channels between any pair of the TX and RX sub-arrays are jointly correlated due to the shared common scatterers in DPA-MIMO systems.  Therefore, it is highly desirable to exploit both common and innovation sparsity to decrease the pilot overhead and improve the accuracy of the DPA-MIMO channel estimate.

With the obtained channel, hybrid precoding should be performed to facilitate directional data transmission in DPA-MIMO systems. Some works have been devoted to hybrid precoding in energy-efficient sub-array architectures~\cite{singh2015feasibility,park2017dynamic,yu2016alternating,gao2016energy}. In~\cite{yu2016alternating}, \textit{Yu et al.} proposed an alternating optimization based method to minimize the Euclidean distance between the fully digital precoder and the hybrid precoders. Therein, during each iteration, the optimal digital precoder is obtained by solving a semidefinite relaxation (SDR) problem with a heavy computational burden while the optimal RF precoder has a closed-form expression.  
The work in~\cite{gao2016energy} designed hybrid precoders by creatively introducing the mechanism of successive interference cancellation (SIC) in multi-user detection. Thus, a total achievable rate maximization problem, with nonconvex constant amplitude constraints of phase shifters, is decomposed  into a series of subrate optimization problems each of which handles one sub-array. 
However, this algorithm only focuses on the TX design with a simple diagonal baseband precoding matrix under the case that the number of data streams equals that of RF chains.
Inspired by~\cite{gao2016energy}, we extend the SIC idea to hybrid precoding in DPA-MIMO systems.

In this paper, we consider cooperative multi-sub-array based channel estimation and hybrid precoding for DPA-MIMO systems. The main contributions are summarized below.  
\begin{itemize}
\item We exploit joint channel sparsity among distributed sub-arrays. The inter-sub-array coupling channel model motivates us to take advantage of a multi-sub-array coordinated channel sounding strategy which undoubtedly decreases the training overhead. Based on this strategy, we formulate the DPA-MIMO channel estimation problem as a structured single measurement vector (SMV) recovery problem in compressed sensing (CS)~\cite{baraniuk2010model}. Instead of using traditional random pilots at the cost of complicated RF hardware, we design deterministic pilot beam patterns by minimizing the total coherence of the equivalent measurement matrix~\cite{zelnik2011sensing}, which have successful applications in  fully-connected hybrid MIMO~\cite{lee2016channel,srivastava2019quasi} and lens arrays~\cite{wan2020compressive}.

\item  We propose two customized algorithms to find the optimal sparse channel vector with equi-length structured blocks each of which has both the common and innovation supports. 
Inspired by~\cite{rao2014distributed}, we divide the proposed joint orthogonal matching pursuit (JOMP) algorithm into two intuitive parts, namely the common support identification and the innovation support identification. 
Another one is called the joint SBL (JSBL)-$\ell_2$ algorithm which adapts the SBL framework~\cite{wipf2004sparse,wipf2011latent,chen2016simultaneous} to the structured DPA-MIMO channel estimation problem by capitalizing on a dual-space transform.

\item We propose a low-complexity SIC-based hybrid precoding scheme through sub-array grouping for the array-of-sub-arrays architecture. For the design of the RF beamformers, the idea of SIC is used to decompose the spectral efficiency (SE) maximization problem into several subproblems each of which is only related to one group of sub-arrays, thereby facilitating efficient handling of the sub-arrays group by group.
\end{itemize}

\textbf{Organization}: The rest of the paper is organized as follows. Section II introduces a jointly sparse DPA-MIMO channel model and a scheme of cooperative multi-sub-array beam training. Section III formulates the DPA-MIMO channel estimation problem and proposes the sub-array based pilot beam pattern design. Section IV presents two channel recovery algorithms based on the structured channel sparsity. Section V specifies the SIC-based hybrid precoding design through sub-array grouping, with simulation results analyzed in Section VI. Finally, Section VII concludes this paper.

\textbf{Notation}: bold uppercase $\mathbf{A}$ (bold lowercase $\mathbf{a}$) denotes a matrix (a vector). We denote $\left[ \mathbf{A}\right]_{i,j}$ and $\left[ \mathbf{A}\right]_{:,j}$ as its $\left( i,j \right)$th element and $j$th column, respectively. ${\mathrm{vec}}\left(  \cdot  \right) $ stacks the columns of a matrix into a tall vector, and ${\mathrm{Tr}}\left \{  \cdot  \right\}$ stands for the matrix trace operation.  ${{\mathbf{I}}_N}$ and ${{\mathbf{0}}_{M,N}}$ denote the $N \times N$ dimensional identity matrix and the $M \times N$ dimensional all-zero matrix, respectively. ${\left(  \cdot  \right)^H}$, ${\left(  \cdot  \right)^T}$ , ${\left(  \cdot  \right)^ * }$, ${\left(  \cdot  \right)^{-1} }$ and ${\left(  \cdot  \right)^ {\dagger} }$ stand for the conjugate transpose, transpose, conjugate, inverse and pseudo-inverse, respectively. ${\mathrm{diag}}\left\{ {\mathbf{a}} \right\}$, ${\mathrm{diag}}\left\{ {\mathbf{A}} \right\}$ and ${\mathrm{blkdiag}}\left\{ {\mathbf{A}_1}, \cdots, {\mathbf{A}_N} \right\}$ represent a diagonal matrix with $\mathbf{a}$ along its main diagonal, a vector constructed by the main diagonal of the matrix $\mathbf{A}$, a block diagonal matrix , respectively. $\otimes$ denotes the Kronecker product of two matrices. $\ell_0$ , $\ell_1$ and $\ell_2$ norm of vectors are denoted by $\left \| \cdot  \right \|_0$, $\left \| \cdot  \right \|_1$ and $\left \| \cdot  \right \|_2$, respectively. ${\left\|  \mathbf{A} \right\|_{\mathrm{F}}}$ denotes the Frobenius norm and the mixed $\ell_{1}/\ell_{2}$ norm is defined as ${\left\| {\mathbf{A}} \right\|_{1,2}} \triangleq \sum\nolimits_i {\sqrt {\sum\nolimits_j {{{| {{{\left[ {\mathbf{A}} \right]}_{i,j}}} |}^2}} } } $. ${\mathcal{CN}}\left( \boldsymbol{\mu } , \mathbf{R} \right)$ denotes the complex Gaussian distribution with mean $\boldsymbol{\mu }$ and covariance matrix $\mathbf{R}$. ${\mathbb{E}}\left\{  \cdot  \right\}$ is the expectation operator. Finally, $\backslash$ denotes the set subtraction operation.

\section{System Model}
In this section, we present the jointly sparse DPA-MIMO channel model and the cooperative multi-sub-array beam training design in the DPA-MIMO system. 

\subsection{Joint Channel Sparsity Model}
\begin{figure*}[!t]
\centering
\includegraphics[width=6.5in]{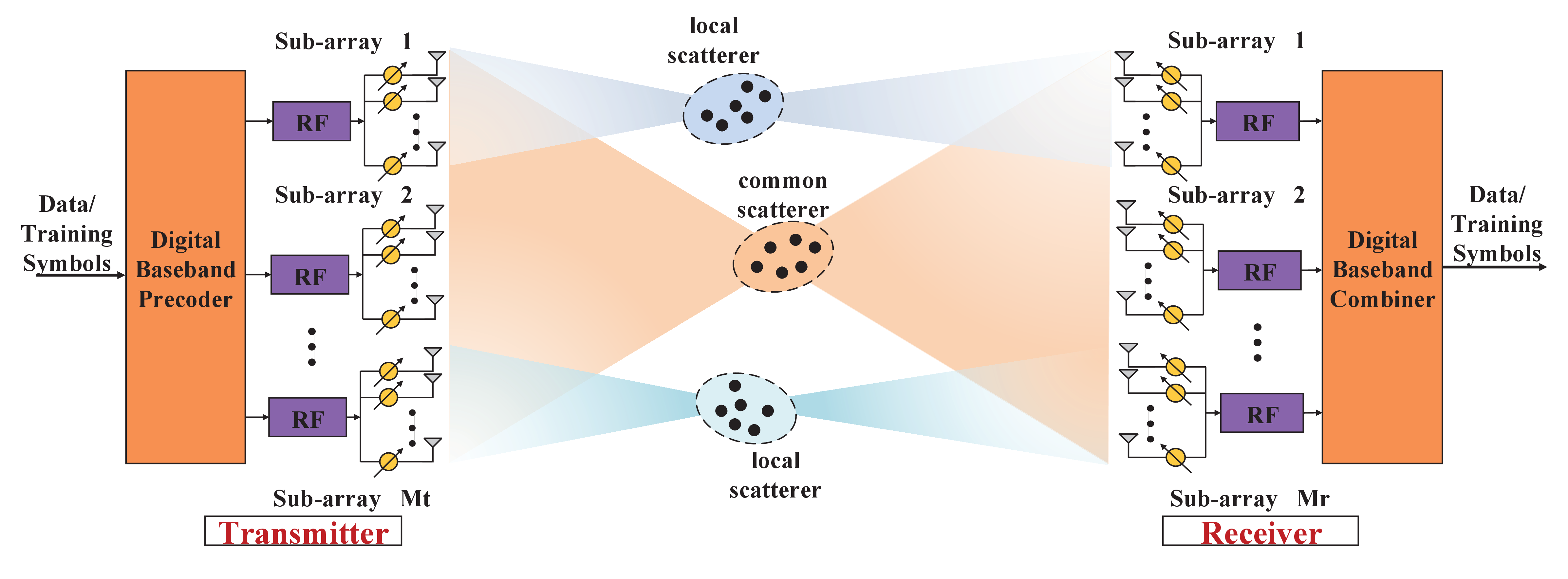}
\caption{An architecture of the DPA-MIMO system and the illustration of the cluster visibility.}
\label{fig_dpa_mimo}
\end{figure*}
\begin{figure}[!htb]
\centering
\includegraphics[width=2.8in]{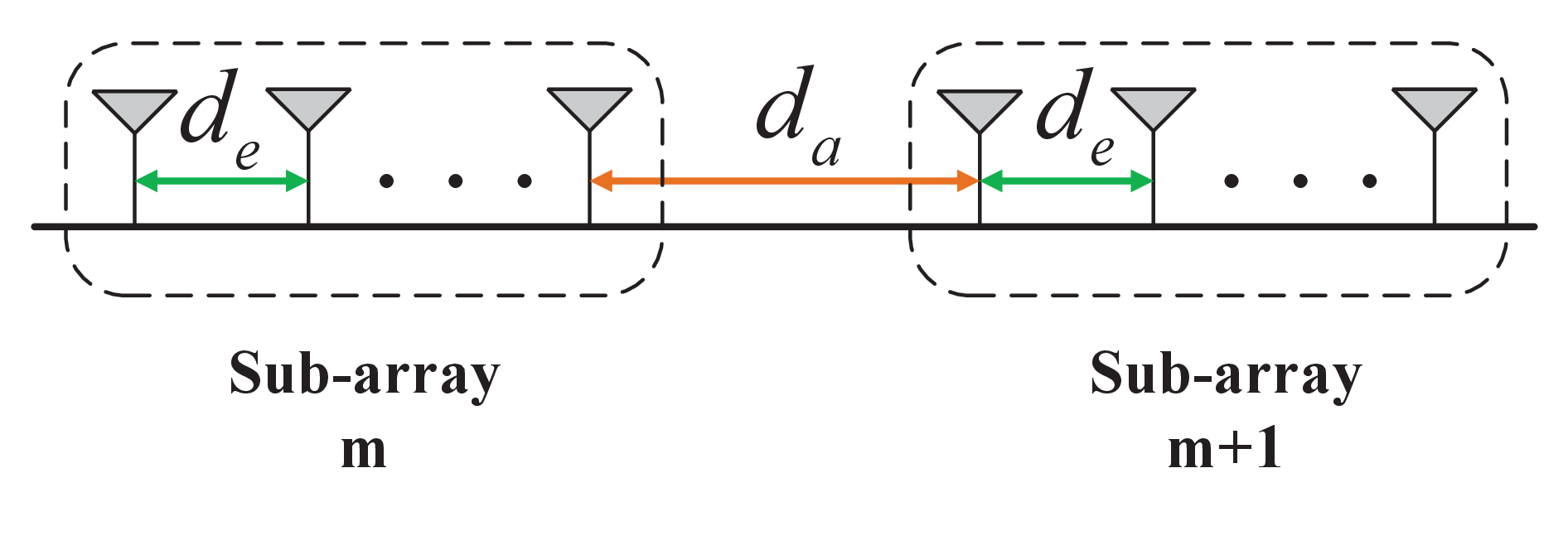}
\caption{Illustration of sub-array spacing}
\label{fig_subarray_spacing}
\end{figure}

Consider a $M_r \times M_t$ DPA-MIMO system shown in Fig.{~\ref{fig_dpa_mimo}}, where a TX with $M_t$ sub-arrays communicates $N_s$ data streams to a RX with $M_r$ subrrays. We denote by $N_t^{\rm tot}$ ($N_r^{\rm tot}$) the total number of antennas at the TX/RX end. Note that $N_t^{\rm tot} = M_t  N_t^{\rm sub}$ and $N_r^{\rm tot} = M_r N_r^{\rm sub}$. Furthermore, we assume that each sub-array is a uniform linear array (ULA){\footnote{The ULA is widely applied to the existing cellular UE and BS, e.g., the performance comparison of single-user precoding schemes with multi-user precoding schemes has been conducted based on ULA configuration in the high-impact work~\cite{raghavan2017single}. Moreover, \cite{raghavan2016beamforming} studied the efficacy of different beamforming approaches for initial UE discovery in mmWave MIMO systems, also based on a typical example of ULA setup case. On the other hand, admittedly, the uniform planar arrays (UPAs) are believed to play a promising and crucial role at both BS~\cite{huo20175g} and UE~\cite{raghavan2019antenna} end, particularly for mmWave cellular communications, since they can accommodate many more antenna elements into a sub-array in two-dimensional (2-D) configuration to enable 3-D beamforming. The proposed DPA-MIMO application can be also extended to the UPA setup by using the Kronecker product based codebook~\cite{xie2013limited}. We expect to extend and discuss this topic carefully in the future work.}}, and all the sub-arrays are equally spaced and are  arranged in the same axis at both TX and RX ends. Fig.{~\ref{fig_subarray_spacing}} shows that $d_e$ is antenna element spacing inside each sub-array and $d_a$ defines the inter-sub-array spacing. In order to avoid causing too serious grating lobes and significant channel capacity degradation in the DPA-MIMO system, the antenna spacing and inter-sub-array spacing usually satisfy $d_e = 0.5\lambda_c $ and $d_a \geq  1.5\lambda_c$ respectively~\cite{huo20175g}. Furthermore, $d_a >> d_e$ can be easily satisfied at mmWave frequencies. For example,  by adopting the DPA-MIMO architecture on a unmanned aerial vehicles (UAV) where $d_a$ is larger than 20 times the free-space wavelength, or 40 times the $d_e$, Gbps data-rate communication for multi-user scenarios without interference can be enabled and has been preliminarily verified from field tests~\cite{huo2019multi,huo2019distributed}.

Compared with the rich scattering channel model often used for microwave frequencies, mmWave channels are better characterized by a limited number of scattering clusters~\cite{raghavan2016beamforming}. Thus, the $L_{m,n}$-path narrowband channel matrix between the $n$th TX sub-array and the $m$th RX sub-array is formulated as
\begin{equation}
\label{equ_channel_subarray}
{{\bf{H}}_{m,n}} = \sqrt {\frac{{N_t^{{\rm{sub}}}N_r^{\rm{sub}}}}{L_{m,n}}} 
\sum\limits_{i = 0}^{{L_{m,n}} - 1} {\alpha _{m,n}^{\left( i \right)}{{\bf{a}}_r}\left( {\vartheta _{m,n}^{\left( i \right)}} \right){\bf{a}}_t^H\left( {\psi _{m,n}^{\left( i \right)}} \right)} ,
\end{equation}
where $\alpha _{m,n}^{\left (0  \right )}$ is the complex gain of the line-of-sight (LoS) component with $\vartheta _{m,n}^{\left( 0 \right)}$ and $\psi _{m,n}^{\left( 0 \right)}$ representing its spatial directions composed of an angle of arrival (AoA) and an angle of departure (AoD), respectively. For $i = 1,2, \cdots, L_{m,n} - 1$, $\alpha _{m,n}^{\left (i  \right )}$ is the complex gain of the $i$th non-LoS (NLoS) component with $\vartheta _{m,n}^{\left( i \right)}$ and $\psi _{m,n}^{\left( i \right)}$ denoting its spatial directions composed of an AoA and an AoD, respectively. Note that some AoDs/AoAs are correlated along different sub-arrays, which will be investigated in the proposed jointly sparse DPA-MIMO channel model defined in Definition {\ref{def_dpa_mimo_chl}}. The path amplitudes are assumed to be Rayleigh distributed, i.e., $\alpha _{m,n}^{\left (0  \right )} \sim \mathcal{CN} \left ( 0, \sigma_{\mathrm{LoS}}^{2} \right )$ and $\alpha _{m,n}^{\left (i  \right )} \sim \mathcal{CN} \left ( 0, \sigma_{\mathrm{NLoS}}^{2} \right )$, where $\sigma_{\mathrm{LoS}}^{2}$ and $\sigma_{\mathrm{NLoS}}^{2}$ are the variances of the LoS and NLoS path gain, respectively~\cite{gao2017reliable}.  For an ULA with $N$ antennas, the array response vector is  
${\mathbf{a}}\left( \psi  \right) \triangleq \frac{1}{{\sqrt N }}{\left[ {1,{e^{ - j\frac{{2\pi {d_e}}}{{{\lambda _c}}}\psi }}, \cdots ,{e^{ - j\frac{{2\pi {d_e}}}{{{\lambda _c}}}\left( {N - 1} \right)\psi }}} \right]^T}$
and the spatial direction is defined as $\psi  = \cos \theta $ where $\theta$ is the physical direction and $\lambda_c$ is the carrier wavelength. We use $\mathbf{a}_{t}\left ( \cdot  \right )$ and $\mathbf{a}_{r}\left ( \cdot  \right )$ to denote the array response vectors for the TX and RX sub-arrays, respectively, and define the entire channel matrix between all TX and RX sub-arrays using $\mathbf{H}$.

\begin{figure}[!ht]
\centering
\includegraphics[width=2.5in]{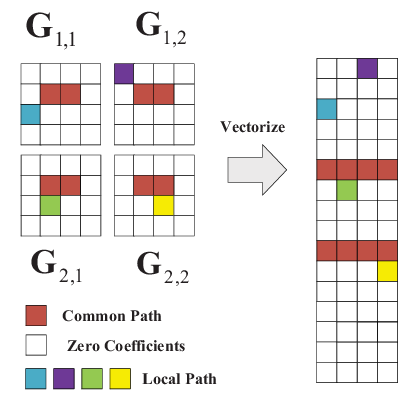}
\caption{Beam-domain representation of the jointly sparse DPA-MIMO channel due to common and local scattering.}
\label{fig_beam_channel}
\end{figure}

In a DPA-MIMO system, since the AoAs or AoDs for different sub-arrays are partially overlapped as shown in Fig.~1, the spatial resolution is actually determined by the number of sub-array antennas. Assuming in an extreme case that two sub-arrays have different AoAs, the dimension of the channel should equal the number of sub-array antennas~\cite{lin2017subarray}. The physical spatial domain and the beam domain are related through a spatial unitary transform matrix, which contains the array steering vectors of uniformly spaced orthogonal spatial directions covering the entire space, e.g., 
$\mathbf{U} = \left [ \mathbf{a}\left ( \bar{\psi}_{1} \right ),\cdots ,\mathbf{a}\left ( \bar{\psi}_{N_G} \right ) \right ]$,
where $\bar{\psi}_{g} = \frac{2}{{{N_G}}}\left( {g - 1} \right) - 1$ for $g=1,\cdots ,N_G$ with $N_G \in \left\{ {N_t^{{\text{sub}}},N_r^{{\text{sub}}}} \right\}$~\cite{sayeed2002deconstructing}. Note that finite-resolution discrete dictionary may cause the power leakage problem. One way to handle this problem is to use a redundant dictionary matrix. Interested readers can refer to~\cite{wan2020compressive} for detailed operations.  We use $\mathbf{U}_{t}$ ($\mathbf{U}_{r}$) to denote the spatial transform matrix for each TX/RX sub-array. Thus, the beam-domain channel matrix between the $n$th TX sub-array and the $m$th RX sub-array can be represented as $\mathbf{H}_{m,n}=\mathbf{U}_{r} \mathbf{G}_{m,n} \mathbf{U}_{t}^{H}$. Subsequently, we can express the relationship between the entire spatial channel and the entire beam-domain channel for the DPA-MIMO system as
${\bf{H}} = \mathbf{A}_r  {\bf{G}} \mathbf{A}_t^{H}$,
where $ \mathbf{A}_t = {{{\bf{I}}_{{M_t}}} \otimes {\bf{U}}_t} $ and $\mathbf{A}_r =  {{{\bf{I}}_{{M_r}}} \otimes {{\bf{U}}_r}} $ constitute the beam-domain transform matrix for the TX and the RX, respectively, and $\mathbf{G}$ is the entire beam-domain channel matrix that has the following form
\begin{equation}
\mathbf{G} = \begin{bmatrix}
\mathbf{G}_{1,1} &  \cdots & \mathbf{G}_{1,M_t} \\ 
\vdots  &  \ddots & \vdots \\ 
\mathbf{G}_{M_r,1}  & \cdots  & \mathbf{G}_{M_r,M_t} 
\end{bmatrix}.
\end{equation}
Since spatially correlated mmWave channels are expected to have limited scattering, the beam-domain subchannels $\mathbf{G}_{m,n}$'s are sparse by neglecting the subtle grid quantization error of the AoAs and AoDs~\cite{sayeed2002deconstructing,alkhateeb2014channel}. Furthermore, the entire beam-domain channel $\mathbf{G}$ is composed of all the subchannels $\mathbf{G}_{m,n}$'s leading to its sparsity equal to the sum of all the components, which is vividly shown in the left part of Fig.~\ref{fig_beam_channel}. In order to decrease the training overhead in the DPA-MIMO system working on mmWave bands, it is more efficient to estimate the sparse beam-domain channel $\mathbf{G}$ instead of the original physical channel $\mathbf{H}$~\cite{gao2017reliable,li2018mmwave}.

As illustrated in Fig.{~\ref{fig_dpa_mimo}}, there exist both the common and local scatterers between any pair of the TX and RX sub-arrays due to large sub-array spacing. In this figure, the VR of the common scatterer covers all the sub-arrays while the VR of any local scatterer only illuminates certain sub-array at the TX/RX end. This result is analogous to that of the newly studied extra-large antenna arrays~\cite{gao2015massive,ali2019linear,de2019non}. In the DPA-MIMO architecture, the structure composed of all sub-arrays at the TX/RX end can be treated as another special version of an extra-large ULA where segments of antenna elements are removed at uniform intervals. Therefore, the spatial non-stationarity of DPA-MIMO means that different scatterers may be observed by different sub-arrays, which is parallel to the spatially non-WSS assumption made on the large array~\cite{zhang2018recent}.
Due to the inherent channel sparsity at mmWave frequencies, it is more convenient to investigate this kind of spatially non-WSS channels from the perspective of virtual angular domain~\cite{rao2014distributed}.
Therefore, we conclude the following assumption on the beam-domain channel matrices in the DPA-MIMO system:
\begin{defn}[\textbf{Jointly Sparse DPA-MIMO Channel}]
\label{def_dpa_mimo_chl}
The channel matrices $ \left \{ \mathbf{G}_{m,n} \right \}$ have the following properties:
\begin{itemize}[leftmargin=*]
\item \textbf{Common sparsity} due to both the LoS path and common scattering:
Denote $\mathrm{supp}\left \{ \mathbf{A}  \right \}$ as the index set of non-zero entries of the matrix $\mathbf{A}$. Then, $ \left \{ \mathbf{G}_{m,n} \right \}$  are simultaneously sparse. Different $ \left \{ \mathbf{G}_{m,n} \right \}$ share a common support, i.e., the index set $\Omega _c$ which satisfies 
\begin{equation}
\label{equ_omega_c}
\Omega _c \triangleq {\bigcap_{m=1}^{M_r}} {\bigcap_{n=1}^{M_t}} \mathrm{supp}\left \{ \mathbf{G}_{m,n} \right \} .
\end{equation}
\item \textbf{Innovation sparsity} due to local scattering: There exist unique components for $\mathbf{G}_{m,n}$, i.e.,
\begin{equation}
\label{equ_omega_mn}
\Omega _{m,n} \triangleq \mathrm{supp}\left \{ \mathbf{G}_{m,n} \right \} \backslash \Omega_c.
\end{equation}
\end{itemize}
\end{defn}
\begin{rem} 
Common sparsity consists of the paths and sub-rays that lead to the same AoA and AoD for different sub-arrays. The exact components of the common sparsity set depend on the TX and RX positions, sub-array spacings, and the relative locations and orientations of the scattering surfaces.   
Moreover, the property of common sparsity is based on the prerequisite that all the sub-arrays at the TX (RX) are installed in parallel, e.g., all the sub-arrays composed of ULAs at the TX (RX) are deployed in the same line~\cite{lota20175g} or all the sub-arrays composed of UPAs at the TX (RX) are parallelly placed on the same plane~\cite{singh2015feasibility,huo20175g}. Some examples of demonstrations are shown in Fig. 5 and Fig. 13 of~\cite{huo20175g}. Furthermore, take mmWave UPAs as an example, the spacing between sub-arrays is proposed to be larger than 1.5 times the free-space wavelength (one wavelength is 10.7 mm when the carrier frequency is 28~GHz)~\cite{huo20175g}, in order to avoid electromagnetic mutual coupling~\cite{rappaport2019wireless}. Thus, several free-space wavelengths that are equal to several centimeters are still relatively small compared to the transmission distance of the LoS communication path and the transmission distance between any common scatterer and any sub-array. As a result, the LoS path or any path caused by common scattering produces almost same AoA and AoD upon all the sub-arrays based on the far-field approximation~\cite{viswanath2005fundamentals}, which leads to a common support in the beam domain. 
\end{rem}
It is observed from Definition {\ref{def_dpa_mimo_chl}} that the DPA-MIMO channel sparsity support is parametrized by the set $\left \{ \Omega_c,\left \{ \Omega _{m,n}  \right \} \right \} $, where $\Omega _{m,n}$ and $\Omega_c$ determine the innovation sparsity support and the shared common sparsity support, respectively. In the most common case of $\Omega_{m,n} = \varnothing$ for $\forall m,n$, all sub-arrays at the TX/RX have the same AoDs (AoAs)~\cite{lota20175g,singh2015feasibility,gao2016energy}. In another case  of $\Omega _{c} = \varnothing $, independent scatterers are present for each sub-array~\cite{lin2015adaptive}. Furthermore, Fig.{~\ref{fig_beam_channel}} is given for  a better understanding of the structured sparsity in the DPA-MIMO channel. In this example, we assume a $2\times 2$ DPA-MIMO system with $N_t^{\mathrm{sub}} = N_r^{\mathrm{sub}} = 4$, and there are the LoS path and one common cluster among all sub-arrays and one local cluster within each transceiver sub-array pair. By applying the vectorized operation to each beam-domain channel matrix $\mathbf{G}_{m,n}$ and putting the obtained channel vectors together, we can formulate a new structured sparse matrix $\left [ \mathrm{vec}\left ( \mathbf{G}_{1,1} \right ),\mathrm{vec}\left ( \mathbf{G}_{2,1} \right ),\mathrm{vec}\left ( \mathbf{G}_{1,2} \right ),\mathrm{vec}\left ( \mathbf{G}_{2,2} \right ) \right ]$ that is row-sparse plus element-sparse. In the following, for both the LoS path and the shared paths formed by the common clusters, we call them the common paths indiscriminately.

\subsection{Cooperative Multi-Sub-Array Beam Training}
Spatial scanning based beam training approaches are widely adopted for mmWave channel estimation due to its simplicity and high performance~\cite{liu2017millimeter,li2019explore,liu2020millimeter}.
However, for the non-cooperative spatial scanning based training process, each sub-array at the TX should individually  spend $N_{t}^{\mathrm{sub}}N_{r}^{\mathrm{sub}}$ training beams defined in discrete Fourier transform (DFT) based RF codebooks~\cite{singh2015feasibility,zhang2016tracking}. In order to decrease the probing overhead, we adopt a cooperative multi-sub-array beam training scheme{\footnote{The hierarchical codebook based scheme~\cite{alkhateeb2014channel,raghavan2016beamforming,xiao2016hierarchical,lin2017subarray} only requires $SL^2\left \lceil SL/M_r \right \rceil \log_S \left ( N^{\mathrm{tot}} / L \right )$ TX training beams, where $L_{m,n} = L $ for $\forall m,n$, $N_t^{\mathrm{tot}} = N_r^{\mathrm{tot}} = N^{\mathrm{tot}}$ and $S$ is a design parameter that is usually set to be 2. However, this low overhead scheme is not suitable for the partial coupling DPA-MIMO channel characterized in Definition {\ref{def_dpa_mimo_chl}}.}}. For CSI acquisition, the TX uses $N_t^{\mathrm{beam}}$ ($N_t^{\mathrm{beam}} \leq N_t^{\mathrm{tot}}$) pilot beam patterns denoted as $\big \{ \mathbf{f}_{p} \in \mathbb{C}^{N_t^{\mathrm{tot}}\times 1}:{\left \| \mathbf{f}_p \right \|_2^2=1},{p=1,\cdots ,N_t^{\mathrm{beam}}}\big \}$, and the RX adopts $N_r^{\mathrm{beam}}$ ($N_r^{\mathrm{beam}} \leq N_r^{\mathrm{tot}}$) pilot beam patterns denoted as $\big \{ \mathbf{w}_{q} \in \mathbb{C}^{N_r^{\mathrm{tot}}\times 1}:{\left \| \mathbf{w}_q \right \|_2^2=1},{q=1,\cdots ,N_r^{\mathrm{beam}}}\big \}$. 

During the training period, the TX successively sends its training beam patterns $\left \{ \mathbf{f}_p\right \}$ which are received by the RX through its beam patterns $\left \{ \mathbf{w}_q\right \} $. The $\left( {q,p} \right)$th received sample for the $p$th TX beam pattern is given by~\cite{srivastava2019quasi}
\begin{eqnarray} 
\label{equ_y_qp}
{y}_{{q},p}=\mathbf{w}_{{q}}^{H}\mathbf{H}\mathbf{f}_px_p+\mathbf{w}_{{q}}^{H}{\tilde {\mathbf{z}}_p} ,
\end{eqnarray}
where $x_p$ is the transmitted pilot symbol and
$\tilde {\mathbf{z}}_p\in \mathbb{C}^{N_{r}^{\mathrm{tot}}\times 1}$ is a noise vector with $\mathcal{CN}\left ( \mathbf{0},\sigma_{\mathrm{z}}^{2}\mathbf{I }_{N_{r}^{\mathrm{tot}}} \right )$.  Collecting $y_{{q},p}$ for ${q}\in \left \{ 1,\cdots ,N_r^{\mathrm{beam}} \right \}$, we have $\mathbf{y}_p\in \mathbb{C}^{N_{r}^{\mathrm{beam}}\times 1}$ given by 
\begin{eqnarray}
\label{equ_yp}
\mathbf{y}_p=\mathbf{W}^{H}\mathbf{H}\mathbf{f}_{p}x_p+ \mathbf{z}_p ,
\end{eqnarray}
where $\mathbf{W}=\left [\mathbf{w}_1,\cdots ,\mathbf{w}_{{N_{r}^{\mathrm{beam}}}}  \right ]\in \mathbb{C}^{N_{r}^{\mathrm{tot}}\times N_{r}^{\mathrm{beam}}}$ and ${\mathbf{z}}_p = {{\mathbf{W}}^H}{\tilde {\mathbf{z}}_p} \in \mathbb{C}^{N_{r}^{\mathrm{beam}}\times 1}$. To represent the received signals for all TX beam patterns, we collect $\mathbf{y}_p$ for $p\in \left \{ 1,\cdots ,N_{t}^{\mathrm{beam}} \right \}$ yielding
\begin{eqnarray}
\label{equ_Y_X}
\mathbf{Y}=\mathbf{W}^{H}\mathbf{H}\mathbf{F}\mathbf{X}_{\mathrm{p}}+\mathbf{Z} ,
\end{eqnarray}
where the concatenated received signal matrix $\mathbf{Y}=\left [ \mathbf{y}_1,\cdots ,\mathbf{y}_{N_{t}^{\mathrm{beam}}} \right ] \in \mathbb{C}^{N_{r}^{\mathrm{beam}}\times N_{t}^{\mathrm{beam}} }$, the complete TX processing matrix $\mathbf{F}=\left [ \mathbf{f}_{1},\cdots , \mathbf{f}_{N_{t}^{\mathrm{beam}}}\right ]\in \mathbb{C}^{N_{t}^{\mathrm{tot}}\times N_{t}^{\mathrm{beam}}}$ and the concatenated noise matrix $\mathbf{Z} = \left[ {{{\mathbf{z}}_1}, \cdots ,{{\mathbf{z}}_{N_t^{{\text{beam}}}}}} \right]\in \mathbb{C}^{N_{r}^{\mathrm{beam}}\times N_{t}^{\mathrm{beam}}}$. In general, we choose the pilot matrix $\mathbf{X}_{\mathrm{p}}=\sqrt{P_{\mathrm{p}}}\mathbf{I}_{N_{t}^{\mathrm{beam}}}$ where $P_{\mathrm{p}}$ is the pilot power per transmission.

In our array-of-sub-arrays architecture, the TX and RX processing matrices are decomposed as $\mathbf{F} = \mathbf{F}_{\rm{R}}\mathbf{F}_{\rm{D}}$ and $\mathbf{W} = \mathbf{W}_{\rm{R}}\mathbf{W}_{\rm{D}}$. Thus, the multiple measurements in \eqref{equ_Y_X} is expressed as 
\begin{equation}
\label{equ_Y}
{\bf{Y}} = \sqrt {{P_{\rm{p}}}} {\bf{W}}_{{\rm{D}}}^H{\bf{W}}_{{\rm{R}}}^H{\bf{H}}{{\bf{F}}_{{\rm{R}}}}{{\bf{F}}_{{\rm{D}}}} + {\bf{Z}} ,
\end{equation}
where $\mathbf{F}_{\mathrm{R}}\in \mathbb{C}^{N_t^{\mathrm{tot}}\times N_t^{\mathrm{tot}}} $ and $\mathbf{W}_{\mathrm{R}}\in \mathbb{C}^{N_r^{\mathrm{tot}}\times N_r^{\mathrm{tot}}} $ denote RF beamforming matrices at the TX and the RX, respectively, while $\mathbf{F}_{\mathrm{D}}\in \mathbb{C}^{N_t^{\mathrm{tot}}\times N_t^{\mathrm{beam}}}$ and $\mathbf{W}_{\mathrm{D}}\in \mathbb{C}^{N_r^{\mathrm{tot}}\times N_r^{\mathrm{beam}}}$ denote the TX and the RX baseband processing matrices, respectively. 
At the TX,  $\mathbf{F}$ is further partitioned into ${{N_t^{{\text{tot}}}} \mathord{\left/
 {\vphantom {{N_t^{{\text{tot}}}} {{M_t}}}} \right.
 \kern-\nulldelimiterspace} {{M_t}}} = N_t^{{\text{sub}}}$ blocks, that is,
\begin{equation}
\label{equ_F_RF_D}
\begin{aligned}
{\mathbf{F}} = \; &\left[ {{{\mathbf{F}}_{{\text{R}},1}}{{\mathbf{F}}_{{\text{D}},1}}, \cdots ,{{\mathbf{F}}_{{\text{R}},{b_1}}}{{\mathbf{F}}_{{\text{D}},{b_1}}}, \cdots ,{{\mathbf{F}}_{{\text{R}},N_t^{{\text{sub}}}}}{{\mathbf{F}}_{{\text{D}},N_t^{{\text{sub}}}}}} \right] \\
 = \;& \underbrace {\left[ {{{\mathbf{F}}_{{\text{R}},1}}, \cdots ,{{\mathbf{F}}_{{\text{R}},{b_1}}}, \cdots ,{{\mathbf{F}}_{{\text{R}},N_t^{{\text{sub}}}}}} \right]}_{{{\mathbf{F}}_{\text{R}}}} \\
& \cdot \underbrace {{\text{blkdiag}}\left\{ {{{\mathbf{F}}_{{\text{D}},1}}, \cdots ,{{\mathbf{F}}_{{\text{D}},{b_1}}}, \cdots ,{{\mathbf{F}}_{{\text{D}},N_t^{{\text{sub}}}}}} \right\}}_{{{\mathbf{F}}_{\text{D}}}} ,
\end{aligned}
\end{equation}
where $\mathbf{F} _{\mathrm{R},b_1} \in \mathbb{C}^{N_{t}^{\mathrm{tot}} \times M_t}$ represents the shared analog processing matrix for all the beam patterns at block $b_1$ and each column of $\mathbf{F}_{\mathrm{D},b_1}\in \mathbb{C}^{M_t\times \frac{N_{t}^{\mathrm{beam}}}{N_{t}^{\mathrm{sub}}}}$ denotes one digital pilot beamforming vector for the corresponding beam pattern at block $b_1$~\cite{lee2016channel,srivastava2019quasi,wan2020compressive}. 
In the same way, the RX admits the following processing by blocks
\begin{subequations}
\label{equ_W_RF_D}
\begin{align}
{{\mathbf{W}}_{\text{R}}}{\text{ }} & = \left[ {{{\mathbf{W}}_{{\text{R}},1}}, \cdots ,{{\mathbf{W}}_{{\text{R}},{b_2}}}, \cdots ,{{\mathbf{W}}_{{\text{R}},N_r^{{\text{sub}}}}}} \right] , \\
{{\mathbf{W}}_{\text{D}}}{\text{ }} & = {\text{blkdiag}}\left\{ {{{\mathbf{W}}_{{\text{D}},1}}, \cdots ,{{\mathbf{W}}_{{\text{D}},{b_2}}}, \cdots ,{{\mathbf{W}}_{{\text{D}},N_r^{{\text{sub}}}}}} \right\},
\end{align}
\end{subequations}
where $\mathbf{W} _{\mathrm{R},b_2} \in \mathbb{C}^{N_{r}^{\mathrm{tot}} \times M_r}$ and $\mathbf{W}_{\mathrm{D},b_2}\in \mathbb{C}^{M_r\times \frac{N_{r}^{\mathrm{beam}}}{N_{r}^{\mathrm{sub}}}}$.
Moreover, each column of the RF processing matrix is zero except for a continuous block of non-zero entries (consisting of the beamforming weights used on the corresponding sub-array), i.e., ${{\mathbf{F}}_{{\text{R}},{b_1}}} = {\text{blkdiag}}\left\{ {{\mathbf{f}}_{{\text{R}},{b_1}}^{\left[ 1 \right]}, \cdots ,{\mathbf{f}}_{{\text{R}},{b_1}}^{\left[ n \right]}, \cdots ,{\mathbf{f}}_{{\text{R}},{b_1}}^{\left[ {{M_t}} \right]}} \right\}$ and ${{\mathbf{W}}_{{\text{R}},{b_2}}} = {\text{blkdiag}}\left\{ {{\mathbf{w}}_{{\text{R}},{b_2}}^{\left[ 1 \right]}, \cdots {\mathbf{w}}_{{\text{R}},{b_2}}^{\left[ m \right]}, \cdots ,{\mathbf{w}}_{{\text{R}},{b_2}}^{\left[ {{M_r}} \right]}} \right\}$, where ${\mathbf{f}_{\mathrm{R},b_1}^{\left [ n \right ]}}  \in \mathbb{C}^{N_t^{\mathrm{sub}}\times 1}$ and ${\mathbf{w}_{\mathrm{R},b_2}^{\left [ m \right ]}} \in \mathbb{C}^{N_r^{\mathrm{sub}}\times 1}$ denote the RF beamforming weights for the $n$th TX sub-array and the $m$th RX sub-array, respectively. Furthermore,  the amplitude of each element of ${\mathbf{f}_{\mathrm{R},b_1}^{\left [ n \right ]}}$ and ${\mathbf{w}_{\mathrm{R},b_2}^{\left [ m \right ]}}$ equals $\frac{1}{{\sqrt {N_t^{{\text{sub}}}} }}$ and $\frac{1}{{\sqrt {N_r^{{\text{sub}}}} }}$, respectively. 

\section{Formulation of Channel Estimation Problem and Pilot Beam Pattern Design}
In this section, we first exploit the jointly sparse nature of the DPA-MIMO channel, and
formulate its channel estimation problem as a structured sparse vector recovery problem. Then, we propose a deterministic beam training scheme.

\subsection{Formulation of Channel Estimation Problem}
To exploit the sparse nature of the DPA-MIMO channel, it is necessary to vectorize the received signal matrix $\mathbf{Y}$ in {\eqref{equ_Y}}. After denoting $\frac{{{\text{vec}}\left( {\mathbf{Y}} \right)}}{{\sqrt {{P_{\text{p}}}} }}$ by $\mathbf{y}\in \mathbb{C}^{N_{t}^{\mathrm{beam}}N_{r}^{\mathrm{beam}}  \times 1}$, we have
\begin{equation}
\begin{aligned}
\label{equ_vec_Y}
\mathbf{y} & \overset{\left ( a \right )}{=}  \left ( \left ( {\bf{F}}_{{\rm{D}}}^{T} {\bf{F}}_{{\rm{R}}}^{T} \right ) \otimes \left ( {\bf{W}}_{{\rm{D}}}^{H} {\bf{W}}_{{\rm{R}}}^{H} \right ) \right ) \mathrm{vec} \left ({\mathbf{H}}  \right ) + \mathbf{z} \\
& \overset{\left ( b \right )}{=} \mathbf{Q}  \mathrm{vec} \left ({\mathbf{G}}  \right ) + \mathbf{z} ,
\end{aligned}
\end{equation}
where (a) follows from the equivalent noise vector $\mathbf{z} \triangleq \frac{1}{{\sqrt {{P_{\text{p}}}} }}\left [ \mathbf{z}_{1}^{T},\cdots , \mathbf{z}_{N_{t}^{\mathrm{beam}}}^{T} \right ]^{T}\in \mathbb{C}^{N_{t}^{\mathrm{beam}}N_{r}^{\mathrm{beam}}  \times 1}$ 
and the properties of Kronecker product, $\mathrm{vec}\left ( \mathbf{ABC} \right ) = \left ( \mathbf{C}^{T} \otimes \mathbf{A}\right )\mathrm{vec}\left ( \mathbf{B} \right )$ and $\left (\mathbf{A} \otimes \mathbf{B}  \right )^{T} = \mathbf{A}^{T} \otimes \mathbf{B}^{T}$~\cite{kaare2000cookbook}, and (b) follows from $\mathrm{vec}\left ( \mathbf{H} \right ) = \left ( \mathbf{A}_t^{\ast } \otimes \mathbf{A}_r \right ) \mathrm{vec}\left ( \mathbf{G} \right )$ and $\left (\mathbf{A} \otimes \mathbf{B}  \right ) \left (\mathbf{C} \otimes \mathbf{D}  \right ) = \left (\mathbf{AC} \right ) \otimes\left ( \mathbf{BD}     \right )$. The equivalent sensing matrix $\mathbf{Q} \in \mathbb{C}^{N_t^{\mathrm{beam}} N_r^{\mathrm{beam}} \times N_t^{\mathrm{tot}} N_r^{\mathrm{tot}}}$ can be defined as 
\begin{equation}
\mathbf{Q} \triangleq \left ( {\bf{F}}_{{\rm{D}}}^{T} {\bf{F}}_{{\rm{R}}}^{T} \mathbf{A}_{t}^{*}\right ) \otimes \left ( {\bf{W}}_{{\rm{D}}}^{H} {\bf{W}}_{{\rm{R}}}^{H} \mathbf{A}_{r} \right ).
\end{equation}

The formulation of the vectorized received signal in (\ref{equ_vec_Y}) represents a sparse formulation of the channel estimation problem as $\mathrm{vec}\left ( \mathbf{G} \right )$ has only $N_{0} = \left | \Omega _c \right | + {\sum\nolimits_{m=1}^{M_r} \sum\nolimits_{n=1}^{M_t}{ \left | \Omega _{m,n}\right |}}$ non-zero elements and $N_{0} \ll N_{t}^{\mathrm{tot}}N_{r}^{\mathrm{tot}}$. This implies that the number of required measurements $N_t^{\mathrm{beam}} N_r^{\mathrm{beam}}$ to detect the non-zero elements can be much less than $N_t^{\mathrm{tot}} N_r^{\mathrm{tot}}$. We expect to exploit the hidden joint sparsity in the beam-domain channel to reduce the required training and improve the performance of channel estimation. For convenience, we exchange the order of elements in $\mathrm{vec} \left ({\mathbf{G}}  \right )$ to get a new vector with $M_t M_r$ equi-length blocks as
\begin{equation}
\label{equ_new_vecG}
\begin{aligned}
{\mathbf{x}} \triangleq \;&  \Big[ \mathbf{x}_{1}^{T},\cdots , \mathbf{x}_{M_t M_r}^{T}  \Big]^{T} \\
= \; & \Big[ \underbrace{\mathrm{vec}^{T} \left ( \mathbf{G}_{1,1} \right )}_{1 \mathrm{th} \; \mathrm{block} }, \cdots , \underbrace{\mathrm{vec}^{T} \left ( \mathbf{G}_{M_r,1} \right )}_{M_r \mathrm{th} \; \mathrm{block}}, \cdots , \\
& \underbrace{ \mathrm{vec}^{T} \left ( \mathbf{G}_{1,M_t} \right )}_{\left ( M_r \left ( M_t - 1 \right )+1 \right) \mathrm{th} \; \mathrm{block}}, \cdots , \underbrace{ \mathrm{vec}^{T} \left ( \mathbf{G}_{M_r,M_t} \right )}_{M_t M_r \mathrm{th} \; \mathrm{block}} \Big]^{T} ,
\end{aligned}
\end{equation}
where the block size is $N_{t}^{\mathrm {sub}} N_{r}^{\mathrm {sub}}$. This manipulation is also shown in Fig. {\ref{fig_beam_channel}}. In this case, the corresponding equivalent measurement matrix $\boldsymbol{\Phi} \triangleq \mathbf{Q}\boldsymbol{\Pi}$ is obtained by exchanging the column order of $\mathbf{Q}$ where $\boldsymbol{\Pi}$ is a column permutation matrix, such that $\boldsymbol{\Phi} {\mathbf{x}} =  \mathbf{Q} \mathrm{vec} \left ({\mathbf{G}}  \right )$. As a result, the problem of DPA-MIMO channel recovery at the RX can be formulated as 
\begin{equation}
\begin{aligned}
 \quad & \underset{\mathbf{x}}{\text{min}}
& & \left \| {\mathbf y} - {\boldsymbol{\Phi}} {\mathbf{x}}  \right \|_2^2 \\
& \;\text{s.t.}
& &  \mathbf{x} \; \textrm{satisfies the joint sparsity model  in Definition~{\ref{def_dpa_mimo_chl}}}.
\end{aligned}
\label{problem_chl_est}
\end{equation}
However, problem \eqref{problem_chl_est} is very challenging due to the common and innovation sparsity requirement in the constraint which is quite different from the conventional CS-recovery problem with a simple sparsity ($\ell_0$-norm) constraint. In addition, the equivalent measurement matrix has to be carefully designed to guarantee the recovery of the non-zero elements of the vector with high probability by using a small number of measurements. 

\subsection{Open-Loop Pilot Beam Pattern Design}
Instead of randomized sensing matrices frequently used for CS-based channel estimation, a deterministic measurement matrix designed by minimizing its total coherence can improve the recovery performance~\cite{zelnik2011sensing,gao2018compressive}. This strategy has been recently applied to pilot beam pattern design for the fully-connected structure~\cite{lee2016channel,srivastava2019quasi}. Due to its excellent performance improvement and zero feedback overhead, this strategy is further applied to channel sounding in the DPA-MIMO system.

Since the total coherence of $\boldsymbol{\Phi} $ is defined by $\mu ^{\mathrm{tot}}\left ( \boldsymbol{\Phi}  \right ) \triangleq \sum_{k=1}^{N_{t}^{\mathrm{tot}}}\sum_{l\neq k}^{N_{t}^{\mathrm{tot}}}\left ( \left [\boldsymbol{\Phi }  \right ]_{:,k}^{H} \left [\boldsymbol{\Phi }  \right ]_{:,l} \right )^2$, we have 
\begin{equation}
\label{equ_mu_tot_decomp}
\begin{aligned}
\mu ^{\mathrm{tot}}\left ( \boldsymbol{\Phi}  \right ) & \overset{\left ( a \right )}{=} \mu ^{\mathrm{tot}}\left ( \mathbf{Q}  \right ) \\
& \overset{\left ( b \right )}{ \leq} \mu ^{\mathrm{tot}} \left ( {\bf{F}}_{{\rm{D}}}^{T} {\bf{F}}_{{\rm{R}}}^{T}  \mathbf{A}_t^{*} \right) \cdot \mu ^{\mathrm{tot}} \left ( {\bf{W}}_{{\rm{D}}}^{H} {\bf{W}}_{{\rm{R}}}^{H} \mathbf{A}_r \right ) , 
\end{aligned}
\end{equation}
where $\left (a  \right )$ follows from the definition of the total coherence, and $\left (b  \right )$ can be derived in a similar way as~\cite[Lemma 7]{lee2016channel} and this upper-bound plays an important role in the decoupling of pilot beam design at both ends.
Therefore, we can decompose the design problem of minimizing $\mu ^{\mathrm{tot}}\left ( \boldsymbol{\Phi}  \right )$ into two separate designs, namely the design of $\mathbf{F}_{\mathrm{D}}$ and $\mathbf{F}_{\mathrm{R}}$ via minimizing $\mu ^{\mathrm{tot}} \left ( {\bf{F}}_{{\rm{D}}}^{T} {\bf{F}}_{{\rm{R}}}^{T}  \mathbf{A}_t^{*} \right) $ and the design of $\mathbf{W}_{\mathrm{D}}$ and $\mathbf{W}_{\mathrm{R}}$ via minimizing $\mu ^{\mathrm{tot}} \left ( {\bf{W}}_{{\rm{D}}}^{H} {\bf{W}}_{{\rm{R}}}^{H} \mathbf{A}_r \right )$. 

Sequentially, we can simplify $\mu ^{\mathrm{tot}} \left ( {\bf{F}}_{{\rm{D}}}^{T} {\bf{F}}_{{\rm{R}}}^{T}  \mathbf{A}_t^{*} \right) $ as
\begin{equation}
\begin{split}
\mu ^{\mathrm{tot}} \left ( {\bf{F}}_{{\rm{D}}}^{T} {\bf{F}}_{{\rm{R}}}^{T}  \mathbf{A}_t^{*} \right)   & \overset{\left ( a \right )}{=} \left \| \left (  {\bf{F}}_{{\rm{D}}}^{T} {\bf{F}}_{{\rm{R}}}^{T}  \mathbf{A}_t^{*}  \right )^H {\bf{F}}_{{\rm{D}}}^{T} {\bf{F}}_{{\rm{R}}}^{T}  \mathbf{A}_t^{*} -\mathbf{I}_{N_t^{\mathrm{tot}}} \right \|_F^2 \\
& \overset{\left ( b \right )}{=}  \left \|  {\bf{F}}_{{\rm{D}}}^{T} {\bf{F}}_{{\rm{R}}}^{T}  \mathbf{A}_t^{*} \mathbf{A}_t^{T}{\bf{F}}_{{\rm{R}}}^{*}{\bf{F}}_{{\rm{D}}}^{*} - \mathbf{I}_{N_t^{\mathrm{beam}}} \right \|_F^2 + d_0\\
& \overset{\left ( c \right )}{=}  \left \|  {\bf{F}}_{{\rm{D}}}^{T} {\bf{F}}_{{\rm{D}}}^{*} - \mathbf{I}_{N_t^{\mathrm{beam}}} \right \|_F^2 + d_0,
\end{split}
\end{equation}
where $d_0 =  N_t^{\mathrm{tot}} - N_t^{\mathrm{beam}} $, $\left ( a \right )$ is to make the equivalent measurement matrix approximate an identity matrix~\cite{zelnik2011sensing}, $\left ( b \right )$ results from the relationship between the Frobenius norm and the trace,  and $\left ( c \right )$ comes from $\mathbf{A}_t \mathbf{A}_t^H = \mathbf{I}_{N_t^{\mathrm{tot}}}$ and $\mathbf{F}_{\text{R}}^T\mathbf{F}_{\text{R}}^*=\mathbf{I}_{N_t^{\text{tot}}}$. 
Thus, we can optimize $\mathbf{F}_{\mathrm{D}}$ by minimizing $\left \|  {\bf{F}}_{{\rm{D}}}^{T} {\bf{F}}_{{\rm{D}}}^{*} - \mathbf{I}_{N_t^{\mathrm{beam}}} \right \|_F^2 $, which can be further transformed into several parallel subproblems for ${b_1} = 1,2, \cdots ,N_t^{{\text{sub}}}$
\begin{equation}
\label{problem_pilot_design}
\begin{aligned}
& \underset{ \left\| {{{\left[ {{{\mathbf{F}}_{{\text{D}},{b_1}}}} \right]}_{:,m}}} \right\|_2^2 = 1}{\text{min}}
& & \Big\|  {\bf{F}}_{{\rm{D}},{b_1}}^{T} {\bf{F}}_{{\rm{D}},{b_1}}^{*} - \mathbf{I}_{\frac{{N_t^{{\text{beam}}}}}{N_t^{{\text{sub}}}}} \Big\|_F^2 
\end{aligned}
\end{equation}
Via relaxing the individual power constraints and then using the method of Lagrange multipliers~\cite[Theorem 2]{lee2016channel}, the optimal baseband precoder of the $b_1$th block is given by 
\begin{equation}
\label{equ_FBB_MTC}
\mathbf{F}^{\star}_{\mathrm{D},b_1} = \bar{\mathbf{U}}_{t} \Big[  \mathbf{I}_{ \frac{N_{t}^{\mathrm{beam}}}{N_{t}^{\mathrm{sub}}} }, \mathbf{0} _{\frac{N_{t}^{\mathrm{beam}}}{N_{t}^{\mathrm{sub}}}, M_t - \frac{N_{t}^{\mathrm{beam}}}{N_{t}^{\mathrm{sub}}}}  \Big]^{T} \bar{\mathbf{V}}_{t}^{H} ,
\end{equation}
where $\bar{\mathbf{U}}_{t} \in \mathbb{C}^{ M_t \times M_t}$ and $\bar{\mathbf{V}}_{t} \in \mathbb{C}^{  \frac{N_{t}^{\mathrm{beam}}}{N_{t}^{\mathrm{sub}}} \times  \frac{N_{t}^{\mathrm{beam}}}{N_{t}^{\mathrm{sub}}}}$ are arbitrary unitary matrices, e.g., unitary DFT matrices. 

In order to make RF pilot beams cover a full range of AoDs, we choose the unitary DFT matrix as the solution  of TX sub-array $m$
\begin{equation}
\label{equ_FRF_opt}
\left [ \left (\mathbf{f}^{\left [ m \right ]}_{\mathrm{R},1}  \right )^{\star}, \cdots , \left (\mathbf{f}^{\left [ m \right ]}_{\mathrm{R},{N_{t}^{\mathrm{sub}}}}  \right )^{\star} \right ] =\mathrm{circshift}\left (  \mathbf{F}_{N_{t}^{\mathrm{sub}}}, m-1 \right ) ,
\end{equation}
where $\mathbf{F}_{N}$ denotes the $N$-dimensional unitary DFT matrix and $\mathrm{circshift}\left ( \mathbf{A},m \right )$ represents moving the columns of a matrix $\mathbf{A}$ to the right for $\left( m \right)$ columns in a circular manner. Different from the previous pilot beam pattern design for the fully-connected structure~\cite{lee2016channel,srivastava2019quasi}, the dimension of the DFT matrix equals the number of sub-array antennas in a DPA-MIMO system due to the joint sparse channels. In this way, the entire TX array simultaneously probes different spatial directions using RF beams. 

Similar operation can be applied to the RX, leading to the optimal combiners as 
\begin{equation}
\label{equ_WBB_MTC}
\mathbf{W}^{\star}_{\mathrm{D},b_2} = \bar{\mathbf{U}}_{r} \Big[\mathbf{I}_{ \frac{N_{r}^{\mathrm{beam}}}{N_{r}^{\mathrm{sub}}} }, \mathbf{0} _{\frac{N_{r}^{\mathrm{beam}}}{N_{r}^{\mathrm{sub}}}, M_r - \frac{N_{r}^{\mathrm{beam}}}{N_{r}^{\mathrm{sub}}}} \Big]^{T} \bar{\mathbf{V}}_{r}^{H},
\end{equation}
\begin{equation}
\label{equ_WRF_opt}
\left [ \left (\mathbf{w}^{\left [ n \right ]}_{\mathrm{R},1}  \right )^{\star}, \cdots , \left (\mathbf{w}^{\left [ n \right ]}_{\mathrm{R},{N_{r}^{\mathrm{sub}}}}  \right )^{\star} \right ] =\mathrm{circshift}\left (  \mathbf{F}_{N_{r}^{\mathrm{sub}}}, n-1 \right ) ,
\end{equation}
where $\bar{\mathbf{U}}_{r} \in \mathbb{C}^{ M_r \times M_r}$ and $\bar{\mathbf{V}}_{r} \in \mathbb{C}^{  \frac{N_{r}^{\mathrm{beam}}}{N_{r}^{\mathrm{sub}}} \times  \frac{N_{r}^{\mathrm{beam}}}{N_{r}^{\mathrm{sub}}}}$ are arbitrary unitary matrices. Furthermore, the noise vector after the designed RF and baseband processing remains i.i.d. Gaussian with $\mathcal{CN}\left ( \mathbf{0},\frac{{\sigma _{\text{z}}^2}}{{{P_{\text{p}}}}} \mathbf{I}_{N_t^{\mathrm{beam}}N_r^{\mathrm{beam}}} \right )$ without prewhitening, which provides much convenience for the further design of CSI recovery algorithms in the next section.

\section{DPA-MIMO Channel Estimation Algorithms}
To solve problem \eqref{problem_chl_est}, we present two customized algorithms, i.e., an OMP based greedy algorithm with low complexity and a SBL inspired algorithm with excellent accuracy. {For notational simplicity, we define the number of training overhead $ N \triangleq N_t^{\mathrm{beam}} N_r^{\mathrm{beam}}$, the number of resolvable spatial directions $ B \triangleq N_t^{\mathrm{sub}} N_r^{\mathrm{sub}}$ and $K \triangleq M_t M_r$, respectively. }

\subsection{Proposed JOMP Algorithm}
The details of the proposed JOMP algorithm are described in Algorithm \ref{alg_jomp} that is inspired by~\cite{rao2014distributed}, where not only the per-link channel sparsity but also the joint sparsity structure are exploited to enable a distributed compressive CSI estimation scheme for multi-user massive MIMO systems. First of all, $\delta _{1}$ and $\delta _{2}$ are defined as the predetermined thresholds  to control the residual error of different loops within reasonable ranges. In principle, they are related to the noise level  $\sigma _{\mathrm{z}}^2$ and the measurement vector length $N$.  Meanwhile, $T_{\mathrm{1,JOMP}}$ and $T_{\mathrm{2,JOMP}}$ are the maximal numbers of iterations to guarantee the convergence. It is obvious that the ideal values of the two iteration numbers should be set as $T_{\mathrm{1,JOMP}} =\left | \Omega _c \right |$ and $T_{\mathrm{2,JOMP}} = \sum_{m=1}^{M_r}\sum_{n=1}^{M_t} \left | \Omega_{m,n}\right |$. Although it is difficult to acquire the true common and innovation sparsity levels ($\mathbb{S} \triangleq \left \{ |\Omega_c|,\left \{ |\Omega _{m,n}|  \right \} \right \} $),  their statistic bounds change over a very long time scale and can be easily obtained from long-term stochastic learning and estimation as in~\cite{rao2014distributed}.
Thus, we can choose a lower bound of the common sparsity $|\Omega_c|$ as the iteration number for the common support identification, while rendering an upper bound of the sum of all the individual sparsity as the iteration number for the innovation support identification $|\Omega_{m,n}|$., i.e., $T_{\mathrm{1,JOMP}} \leq \left | \Omega _c \right |$  and $T_{\mathrm{2,JOMP}} \geq \sum_{m=1}^{M_r}\sum_{n=1}^{M_t} \left | \Omega_{m,n}\right |$. 
 
The proposed JOMP algorithm is further divided into two parts, where the first part aims at common support identification, and the second part continues the innovation support identification. Observe that the estimation target $\mathbf{x}$ in {\eqref{equ_new_vecG}} has non-zero elements at the same positions of each block. Therefore, motivated by the simultaneous sparse approximation algorithm proposed for MMV problems in~\cite{tropp2006algorithms}, we wish to find a group of equi-spaced atoms in the equivalent measurement matrix $\boldsymbol{\Phi}$ by maximizing the sum of their absolute correlations with the residual $\mathbf{r}_1$. This procedure is done in step 1 and the absolute sum has the equivalent expression as $\sum_{k=1}^{K}\left \| \boldsymbol{\Phi}_{:,b+\left ( k-1 \right )B}^{H} \mathbf{r}_1  \right \|_2^2 = \Big \| \left (\boldsymbol{\Phi}_{\Omega _{b}^{K}}  \right )^{H} \mathbf{r}_1 \Big \|_{2}^{2}$. After the  common support $\Omega _{c}^{e}$ is detected, the standard OMP method in~\cite{tropp2007signal} is used to identify the innovation support $\Omega _{i}^{e}$ as realized from step 5 to 8. Depending on the estimated support index, the LS method is finally used to recovery the channel vector.
\begin{algorithm}[!ht]
\label{alg_jomp}
\caption{JOMP Algorithm}
\SetKwInOut{Input}{Input}
\SetKwInOut{Output}{Output}
\SetKwFor{While}{while}{}{}%
\Input{ $\boldsymbol{\Phi}$, $\mathbf{y}$, $T_{\mathrm{1,JOMP}}$, $T_{\mathrm{2,JOMP}}$, $\delta _{1}$, $\delta _{2}$ .}
\textbf{Part 1} (\textit {Common Support Identification}): Initialize $\Omega _{c}^{e} = \varnothing$, $\Omega ^{a} = \left \{ 1,\cdots ,BK  \right \}$, $\mathbf{r}_1 = \mathbf{y}$, and  $\Omega _{b}^{K} = \left \{ b,B+b,\cdots ,\left ( K-1 \right )B+b \right \} $ for $1\leq b\leq B$. \\
\While{$t_1\le T_{\mathrm{1,JOMP}}$ $\mathrm{or}$ $\left \| \mathbf{r}_1 \right \|_{2}^{2} > \delta _{1}$}
{
1. (\textit{Support Estimate}): $ b^{\star } = \mathrm{arg \;\underset{1\leq b\leq B}{max} } \; \left \| \left (\boldsymbol{\Phi}_{\Omega _{b}^{K}}  \right )^{H} \mathbf{r}_1 \right \|_{2}^{2}$. \\
2. (\textit{Support Update}): $\Omega _{c}^{e} = \Omega _{c}^{e} \cup \left \{ \Omega _{b^{\star }}^{K} \right \}$. \\
3. (\textit{Residual Update}): $\mathbf{r}_1 = \mathbf{y} - \boldsymbol{\Phi}_{\Omega _{c}^{e}} \left (\boldsymbol{\Phi}_{\Omega _{c}^{e}}  \right )^{\dagger} \mathbf{y} $. \\
4. (\textit{Iteration Update}): $t_1 = t_1+1$.
}
\textbf{Part 2} (\textit {Innovation Support Identification}): Set $\Omega _{i}^{e} = \Omega _{c}^{e}$, $\mathbf{r}_2 = \mathbf{r}_1$ and $\Omega ^{r} = \Omega ^{a} \setminus \Omega _{c} ^{e}$.\\
\While{$t_2\le T_{\mathrm{2,JOMP}}$ $\mathrm{or}$ $\left \| \mathbf{r}_2 \right \|_{2}^{2} > \delta _{2}$}
{
5. (\textit{Support Estimate}): $j^{\star } =\mathrm{arg} \; \underset{j \in \Omega ^{r} }{\max} \; \left | \left [ \boldsymbol{\Phi} \right ] _{:,j}^{H} \mathbf{r}_2  \right |^{2}  $. \\
6. (\textit{Support Update}): $\Omega _{i}^{e} = \Omega _{i}^{e} \cup \left \{ j^{\star } \right \}$. \\
7. (\textit{Residual Update}): $\mathbf{r}_2 = \mathbf{y} - \boldsymbol{\Phi}_{\Omega _{i}^{e}} \left (\boldsymbol{\Phi}_{\Omega _{i}^{e}}  \right )^{\dagger} \mathbf{y} $. \\
8. (\textit{Iteration Update}): $t_2 =  t_2+1$.
}
\Output  {$\hat{\mathbf{x}}_{{\Omega _{i}^{e}}} = \boldsymbol{\Phi}_{\Omega _{i}^{e}} \left (\boldsymbol{\Phi}_{\Omega _{i}^{e}}  \right )^{\dagger} \mathbf{y}$ and $\hat{\mathbf{x}}_{\Omega ^{a} \setminus \Omega _{c} ^{e}} = \mathbf{0}$.}
\end{algorithm}

\subsection{Proposed JSBL-$\ell_2$ Algorithm}
\subsubsection{Introduction to SBL} 
Consider the classical sparse recovery model without the structured sparsity 
$\mathbf{y} = \boldsymbol{\Phi } \mathbf{x} + \mathbf{z}$,
where $\mathbf{z}$ is a noise vector with $ \mathcal{CN}\left ( \mathbf{0}, \lambda \mathbf{I}_{N} \right )$ and $\lambda $ is the known equivalent noise variance. Thus we have the Gaussian likelihood model $p\left ( \mathbf{y}| \mathbf{x}\right ) = \mathcal{CN}\left ( \boldsymbol{\Phi } \mathbf{x},\lambda \mathbf{I}_{N} \right )$. Assume the parametrized Gaussian prior $p\left ( \mathbf{x} \right ) =\mathcal{CN} \left ( \mathbf{0},\boldsymbol{\Gamma}  \right )$, where $\boldsymbol{\Gamma} = \mathrm{diag}\left \{ \boldsymbol{\gamma}  \right \}$ with a vector of hyper-parameters $\boldsymbol{\gamma}$ governing the prior variances of the elements in $\mathbf{x}$. This latent variable based Gaussian assumption is reasonable from the perspective of variational approximations, which is verified in~\cite[Sec. V]{wipf2004sparse}. For a fixed $\boldsymbol{\gamma}$, using the Bayesian rules we can obtain the Gaussian posterior density of $\mathbf{x}$ as $p\left ( \mathbf{x}| \mathbf{y}\right ) = \mathcal{CN}\left ( \boldsymbol{\mu}_x ,\boldsymbol{\Sigma } _x\right )$, where $\boldsymbol{\mu}_x = \boldsymbol{\Gamma} \boldsymbol{\Phi }^{H}\left ( \lambda \mathbf{I}_{N}+ \boldsymbol{\Phi }\boldsymbol{\Gamma} \boldsymbol{\Phi }^{H}  \right )^{-1} \mathbf{y}$ and $\boldsymbol{\Sigma } _x  = \boldsymbol{\Gamma} - \boldsymbol{\Gamma} \boldsymbol{\Phi }^{H} \left ( \lambda \mathbf{I}_{N} + \boldsymbol{\Phi }\boldsymbol{\Gamma} \boldsymbol{\Phi }^{H}  \right )^{-1}\boldsymbol{\Phi }\boldsymbol{\Gamma}$.
The next key task is to estimate the latent variables $\boldsymbol{\gamma}$. By treating $\mathbf{x}$ as the hidden variables and integrating them out~\cite{wipf2004sparse}, we obtain the maximum a posterior (MAP) estimate on $\boldsymbol{\gamma}$ as
\begin{equation}
\label{equ_gamma_II}
\begin{aligned}
\boldsymbol{\gamma}_{\left ( II \right )} &= \text{arg}\;\underset{\boldsymbol{\gamma }\succeq \mathbf{0}}{\text{max}}\; \int p\left ( \mathbf{y}| \mathbf{x}\right ) p\left ( \mathbf{x}; \boldsymbol{\gamma }\right ) d \mathbf{x} \\
&  = \text{arg}\;\underset{\boldsymbol{\gamma }\succeq \mathbf{0}}{\text{min}}\; \mathbf{y}^{H}\boldsymbol{\Sigma}_{y} ^{-1} \mathbf{y}+\ln\left | \boldsymbol{\Sigma}_{y} \right |  ,
\end{aligned}
\end{equation}
where the covariance matrix of $\mathbf{y}$ denotes $\boldsymbol{\Sigma}_{y}  = \lambda \mathbf{I}_N+\boldsymbol{\Phi }\boldsymbol{\Gamma }\boldsymbol{\Phi }^{H}$. Once $\boldsymbol{\gamma}_{\left ( II \right )}$ is obtained, a commonly accepted point estimate for $\mathbf{x}$ naturally emerges as
\begin{equation}
\label{equ_x_II}
\mathbf{x}_{\left ( II \right )}  = \mathbb{E}\left [ \mathbf{x}|\mathbf{y};\boldsymbol{\gamma}_{\left ( II \right )} \right ] \\
= \boldsymbol{\Gamma }_{\left ( II \right )} \boldsymbol{\Phi}^{H} \left ( \lambda \mathbf{I} + \boldsymbol{\Phi} \boldsymbol{\Gamma }_{\left ( II \right )} \boldsymbol{\Phi}^{H} \right )^{-1} \mathbf{y}.
\end{equation}
This procedure is referred to \textsl{Type II} estimation, also called empirical Bayesian. From {\eqref{equ_x_II}}, it can be observed that a sparse $\boldsymbol{\gamma}_{\left ( II \right )}$ leads to a corresponding sparse estimate $\mathbf{x}_{\left ( II \right )}$. Note that the logarithm term $\ln\left | \boldsymbol{\Sigma}_{y} \right |$ in {\eqref{equ_gamma_II}} is a concave function  with respect to $\boldsymbol{\gamma }$ according to~\cite[Lemma 1]{wipf2004sparse}, thereby favoring a sparse $\boldsymbol{\gamma}$, which further results in a sparse $\mathbf{x}$ through {\eqref{equ_x_II}}. 
The traditional SBL algorithm assumes independent priors for an estimated signal, which however fails to consider the joint sparsity property in the DPA-MIMO channel.

\subsubsection{SBL-Inspired Cost Function}
By reshaping the vector $\mathbf{x}$, we define a new matrix $\mathbf{X} \triangleq \left [ \mathbf{x}_{1},\cdots , \mathbf{x}_{K}\right ] \in \mathbb{C}^{B\times K}$ which is both row-sparse and element-sparse, as shown in Fig. {\ref{fig_beam_channel}}. In order to promote such a structure, $\mathbf{X}$ can be viewed as the summation of an element-sparse matrix $\mathbf{S} \triangleq \left [ \mathbf{s}_{1},\cdots , \mathbf{s}_{K}\right ] \in \mathbb{C}^{B\times K}$ and a row-sparse matrix $\mathbf{C} \triangleq \left [ \mathbf{c}_{1},\cdots , \mathbf{c}_{K}\right ] \in \mathbb{C}^{B\times K}$~\cite{chen2016simultaneous}. Furthermore, by using convex approximation, problem \eqref{problem_chl_est} can be transformed into the following convex optimization problem{\footnote{In fact, we just provide one of the intuitive transformations to tackle the challenging problem in \eqref{problem_chl_est}. As SBL-like algorithms adapt to the measurement matrix and promote a sparse solution, we further address the formulated optimization problem from an SBL perspective.  Provided that one finds a special structured prior that contains both common and individual sparsity, turbo-type message passing algorithms may further improve the estimation performance~\cite{huang2018asymptotically,huang2018iterative}. This meaningful topic requires further investigation.}}:
\begin{eqnarray}
\label{equ_problem_convex}
\underset{\mathbf{C},\mathbf{S}}{\mathrm{min}} \left \| \mathbf{y} - \boldsymbol{\Phi}   {\text{vec}}\left( {{\mathbf{C}} + {\mathbf{S}}} \right) \right \|_2^2 + \beta _{1} \sum_{k=1}^{K}\left \| \mathbf{s}_k \right \|_{1} + \beta _{2}\left \| \mathbf{C} \right \|_{1,2}
\end{eqnarray}
where $\beta _{1} \geq 0$ and $\beta _{2} \geq 0$ are weights regarding element-sparsity and row-sparsity respectively. 

In order to promote sparsity of the solution, we transform the cost function of (\ref{equ_problem_convex}) in $x$-space to the SBL-like cost function in $\gamma$-space by using a dual-space view~\cite{wipf2011latent}, where the following variational representations are used~\cite{bach2012optimization} :
\begin{subequations}
\label{equ_x_norm}
\begin{align}
{\left\| {{{\mathbf{x}}_k}} \right\|_1} &= \mathop {{\text{min}}}\limits_{\gamma _{bk}^s \geqslant 0} \frac{1}{2}\sum\limits_{b = 1}^B {\Big( {\frac{{{{\left| {{x_{bk}}} \right|}^2}}}{{\gamma _{bk}^s}} + \gamma _{bk}^s} \Big)}, \\
 {\left\| {\mathbf{X}} \right\|_{1,2}} & = \mathop {{\text{min}}}\limits_{\gamma _b^c \geqslant 0} \frac{1}{2}\sum\limits_{b = 1}^B {\Big( {\frac{{\sum\nolimits_{k = 1}^K {{{\left| {{x_{bk}}} \right|}^2}} }}{{\gamma _b^c}} + \gamma _b^c} \Big)}   ,
\end{align}
\end{subequations}
where $x _{bk} \triangleq \left [  \mathbf{X} \right ]_{b,k}$, $\gamma _{b}^{c}$ and $\gamma _{bk}^{s}$ are scalars, $\boldsymbol{\gamma }^{c} \triangleq \left [ \gamma _{1}^{c},\cdots ,\gamma _{B}^{c}\right ]^{T}$ is a vector common to all columns of $\mathbf{X}$, and $ \boldsymbol{\gamma }^{s} \triangleq \left [ \gamma _{1,1}^{s},\cdots ,\gamma _{B,1}^{s}, \cdots , \gamma _{1,K}^{s},\cdots ,\gamma _{B,K}^{s}\right ]^{T}$ is a vector with each element corresponding to that of $\mathrm{vec}\left ( \mathbf{X} \right )$. By using the identity (its derivation is given in Appendix~\ref{appendix_derivation}) 
\begin{equation}
\label{equ_ySigmay}
\begin{aligned}
\mathbf{y}^{H} \left (\boldsymbol{\Sigma }^{sc} \right )^{-1} \mathbf{y} & = \underset{\mathbf{c},\mathbf{s}}{\mathrm{min}} \; \frac{1}{\lambda} \left \| \mathbf{y} - \boldsymbol{\Phi} \left ( \mathbf{c}+\mathbf{s} \right ) \right \|_{2}^{2} \\
 & + \mathbf{s}^{H} \left ( \boldsymbol{\Gamma }^{s} \right )^{-1} \mathbf{s}  + \mathbf{c}^{H} \left ( \mathbf{I}_{K} \otimes \boldsymbol{\Gamma }^{c} \right )^{-1} \mathbf{c},
\end{aligned}
\end{equation}
where $\mathbf{s} \triangleq \left [ \mathbf{s}_{1}^{T},\cdots , \mathbf{s}_{K}^{T} \right ]^{T} \in \mathbb{C}^{BK \times 1}$, $\mathbf{c} \triangleq \left [ \mathbf{c}_{1}^{T},\cdots , \mathbf{c}_{K}^{T} \right ]^{T} \in \mathbb{C}^{BK \times 1}$, $\boldsymbol{\Gamma  }^{c} \triangleq \mathrm{diag} \left \{ \boldsymbol{\gamma }^{c} \right\}$, $\boldsymbol{\Gamma  }^{s} \triangleq \mathrm{diag} \left \{ \boldsymbol{\gamma }^{s} \right \}$, and 
\begin{equation}
\boldsymbol{\Sigma }^{sc} \triangleq \lambda \mathbf{I}_{N} + \boldsymbol{\Phi} \left ( \boldsymbol{\Gamma }^{s} + \mathbf{I}_{K} \otimes \boldsymbol{\Gamma }^{c} \right ) \boldsymbol{\Phi}^{H},
\end{equation}
we can further express the convex cost function of (\ref{equ_problem_convex}) in $\gamma$-space as
\begin{equation}
\label{equ_L_I}
\mathcal{L}_{\left ( I \right )} \left (  \boldsymbol{\gamma}^c,\boldsymbol{\gamma}^s \right ) =\mathbf{y}^{H} \left (\boldsymbol{\Sigma }^{sc} \right )^{-1} \mathbf{y} + \beta \mathrm{Tr}\left ( \boldsymbol{\Gamma }^{s} \right ) +  \mathrm{Tr}\left ( \mathbf{I}_{K} \otimes \boldsymbol{\Gamma }^{c} \right ).
\end{equation}
Comparing the data-related term $\mathbf{y}^{H} \left (\boldsymbol{\Sigma }^{sc} \right )^{-1} \mathbf{y}$ in {\eqref{equ_L_I}} and that of the SBL cost function in {\eqref{equ_gamma_II}}, we can observe that the common component $\boldsymbol{\gamma }^{c}$ and the innovation component $\boldsymbol{\gamma }^{s}$  interact with each other in a manner like  $\boldsymbol{\Gamma } = \boldsymbol{\Gamma }^{s} + \mathbf{I}_{K} \otimes \boldsymbol{\Gamma }^{c}$.

Following the innovative decoupling idea in~\cite{chen2016simultaneous}, we can replace the convex penalties in the existing models with the SBL counterpoints to obtain some of the corresponding benefits, even without any formal probabilistic model for this derivation{\footnote{As a matter of fact, \cite{chen2016simultaneous} was the first attempt to apply simultaneous SBL approximation to the two multitask structured sparse models with real variables, i.e., row-sparse with embedded element-sparse and row-sparse plus element-sparse. In this paper, we extend the SBL framework to perform the recovery of the complex structured SMV defined in \eqref{equ_new_vecG} for DPA-MIMO channel estimation.}}. Therefore, we put forth a new cost function in $\gamma$-space from \eqref{equ_L_I} as
\begin{equation}
\label{equ_L_II}
\mathcal{L}_{\left ( II \right )} \left ( \boldsymbol{\gamma}^c,\boldsymbol{\gamma}^s \right ) = \mathbf{y}^{H} \left (\boldsymbol{\Sigma }^{sc} \right )^{-1} \mathbf{y} + \beta \ln \left | \boldsymbol{\Sigma}^{s}  \right | + \ln \left | \boldsymbol{\Sigma}^{c}  \right | ,
\end{equation}
where 
\begin{subequations}
\begin{align}
\boldsymbol{\Sigma }^{s} & \triangleq \frac{\lambda }{2} \mathbf{I}_{N} + \boldsymbol{\Phi } \boldsymbol{\Gamma }^{s} \boldsymbol{\Phi }^{H},\\
\boldsymbol{\Sigma }^{c} & \triangleq \frac{\lambda }{2} \mathbf{I}_{N} + \boldsymbol{\Phi } \left ( \mathbf{I}_{K} \otimes \boldsymbol{\Gamma }^{c}  \right ) \boldsymbol{\Phi }^{H}.
\end{align}
\end{subequations}
Since the log-determinant function is concave and non-decreasing, the term $\ln \left | \boldsymbol{\Sigma}^{c}  \right |$ and the term $\ln \left | \boldsymbol{\Sigma}^{s}  \right |$ promote a sparse common component $\boldsymbol{\gamma}^{c}$ and a sparse innovation component $\boldsymbol{\gamma}^{s}$, respectively. Moreover, the weight $\beta$ regarded as a tradeoff between row sparsity and element sparsity in the defined matrix $\mathbf{X}$, should be tuned with training data or given with prior information. Note that this weight depends on the large-scale properties of the scattering environment and changes over a long time scale.

\subsubsection{$\ell_2$ Reweighting Scheme}
Following an extension of the duality space analysis for the basic SBL framework~\cite{wipf2011latent}, we can transform the cost function of {\eqref{equ_L_II}} from $\gamma$-space to $x$-space. First, via using the identity {\eqref{equ_ySigmay}} and standard determinant identities, we can upper-bound {\eqref{equ_L_II}}  by
\begin{equation}
\label{equ_L_upper}
\begin{aligned}
& \mathfrak{L}\left ( \boldsymbol{\gamma}^c,\boldsymbol{\gamma}^s, \mathbf{c}, \mathbf{s} \right ) =   \frac{1}{\lambda} \left \| \mathbf{y} - \boldsymbol{\Phi} \left ( \mathbf{c}+\mathbf{s} \right ) \right \|_{2}^{2} 
 + \beta \ln  \left | {\boldsymbol{\Gamma }^{s}} \right | \\
& + K \ln \left |{ \boldsymbol{\Gamma }^{c}}\right | + BK\left (\beta +1 \right )  \ln \left ( \frac{\lambda }{2} \right ) + \beta h_{s}\left ( \mathbf{z}^{s} \right ) \\
&+h_{c}\left ( \mathbf{z}^{c} \right )  + \sum_{b=1}^{B} \sum_{k=1}^{K}  {\frac{\left | s_{bk} \right |^{2}}{\gamma _{bk}^{s}}} + \sum_{b=1}^{B}\frac{\sum_{k=1}^{K}\left | c_{bk} \right |^{2}}{\gamma _{b}^{c}}  ,
\end{aligned} 
\end{equation}
where we define two concave functions, namely 
\begin{subequations}
\begin{align}
h_{s}\left ( \boldsymbol{\gamma }^s \right ) & \triangleq \ln \left | \left ( \boldsymbol{\Gamma }^s   \right )^{-1} + \frac{2}{\lambda } \boldsymbol{\Phi }^{H}\boldsymbol{\Phi } \right |,\\
h_{c}\left ( \boldsymbol{\gamma }^c \right ) & \triangleq \ln \left | \left ( \mathbf{I}_{K} \otimes \boldsymbol{\Gamma }^{c}  \right )^{-1} + \frac{2}{\lambda } \boldsymbol{\Phi }^{H}\boldsymbol{\Phi }\right |.
\end{align}
\end{subequations}

Due to the duality of concave conjugate functions, we have the following upper bounds given by
\begin{subequations}
\label{equ_zs_zc}
\begin{align}
h_{s}\left ( \boldsymbol{\gamma }^s \right ) &= \underset{\mathbf{z}^{s}\succeq \mathbf{0}}{\mathrm{min}} \sum\limits_{b=1}^{B}\sum\limits_{k=1}^{K} \Big(\frac{z_{bk}^{s}}{\gamma _{bk}^{s}}- \bar{h}_{s}\left ( \mathbf{z}^{s} \right ) \Big) ,\\
h_{c}\left ( \boldsymbol{\gamma }^c \right ) & = \underset{\mathbf{z}^{c}\succeq \mathbf{0}}{\mathrm{min}}\sum\limits_{b=1}^{B} \Big(\frac{z_{b}^{c}}{\gamma _{b}^{c}}- \bar{h}_{c}\left ( \mathbf{z}^{c}\right ) \Big).
\end{align}
\end{subequations} 
By using {\eqref{equ_L_upper}} and {\eqref{equ_zs_zc}}, we can then perform block coordinate descent (BCD) optimization over the following approximation with irrelevant terms dropped:
\begin{equation}
\label{equ_L_approx}
\begin{aligned}
&  \underset{ \mathcal{Z} }{\mathrm{min}} \left \| \mathbf{y} - \boldsymbol{\Phi} \left ( \mathbf{c}+\mathbf{s} \right ) \right \|_{2}^{2}  
 + \lambda \Big [ \sum_{b=1}^{B} \sum_{k=1}^{K}  \Big ( {\frac{\left | s_{bk} \right |^{2} + \beta z_{bk}^{s}}{\gamma _{bk}^{s}}}+ \beta \ln \gamma_{bk}^{s}  \Big )  \\
 & - \beta \bar{h}_{s}\left ( \mathbf{z}^{s} \right )  
 + \sum_{b=1}^{B}\Big (\frac{\sum_{k=1}^{K}\left | c_{bk} \right |^{2} + z_{bk}^{c}}{\gamma _{b}^{c}} +K \ln \gamma_{b}^{c} \Big )  - \bar{h}_{c}\left ( \mathbf{z}^{c} \right ) \Big ]
\end{aligned}
\end{equation}
where $\mathcal{Z} \triangleq  \lbrace\mathbf{c}, \mathbf{s}, \boldsymbol{\gamma}^c,\boldsymbol{\gamma}^s, \mathbf{z}^{c}, \mathbf{z}^{s} \rbrace$.

By fixing the other variables, we first calculate the optimal values of $\mathbf{s}$ and $\mathbf{c}$. In this way, the optimization problem~\eqref{equ_L_approx} is equivalent to problem~\eqref{equ_ySigmay}. The optimal solutions are expressed as
\begin{subequations}
\label{equ_opt_s_c}
\begin{align}
\mathbf{s}^{\star } &= \boldsymbol{\Gamma }^{s} \boldsymbol{\Phi }^{H} \left ( \boldsymbol{\Sigma }^{sc} \right )^{-1} \mathbf{y},\\
 \mathbf{c}^{\star } & = \left( \mathbf{I}_{K} \otimes \boldsymbol{\Gamma }^{c} \right) \boldsymbol{\Phi }^{H} \left ( \boldsymbol{\Sigma }^{sc} \right )^{-1} \mathbf{y}.
\end{align}
\end{subequations}
which are also proved in Appendix~\ref{appendix_derivation}.

We then optimize $\mathbf{z}^c$ and $\mathbf{z}^s$. According to the duality relationship in {\eqref{equ_zs_zc}}, their optimal values are obtained as
\begin{subequations}
\label{equ_opt_zs_zc}
\begin{align}
\left (\mathbf{z}^{s}  \right )^{\star }  = \;& \mathrm{diag} \left \{ \boldsymbol{\Gamma }^s - \boldsymbol{\Gamma }^s \boldsymbol{\Phi }^{H} \left ( \boldsymbol{\Sigma }^{s}\right )^{-1} \boldsymbol{\Phi } \boldsymbol{\Gamma }^s   \right \} , \\
\left (\mathbf{z}^{c}  \right )^{\star }  = \;& \boldsymbol{\Xi }\cdot \mathrm{diag} \Big\{ \mathbf{I}_{K}\otimes\boldsymbol{\Gamma }^c \\
& - \left( \mathbf{I}_{K}\otimes\boldsymbol{\Gamma }^c\right)\boldsymbol{\Phi }^{H} \left ( \boldsymbol{\Sigma }^{c}\right )^{-1}  
  \boldsymbol{\Phi }\left( \mathbf{I}_{K}\otimes\boldsymbol{\Gamma }^c\right)  \Big\} ,
\end{align}
\end{subequations}
where the Moore-Penrose pseudo-inverse~\cite{kaare2000cookbook} is used for computation reduction and $\boldsymbol{\Xi } \triangleq \left [ \mathbf{I}_B,\cdots , \mathbf{I}_B \right ] \in \mathbb{C} ^{B\times BK}$. Sequentially, the optimal hyperparameters are given by
\begin{subequations}
\label{equ_opt_gamma_s_c}
\begin{align}
\left (\gamma _{bk}^{s}  \right )^{\star } &=  \frac{\left | s_{bk} \right |^2}{\beta}  + z_{bk}^{s}, \\
\left (\gamma _{b}^{c}  \right )^{\star } &= \frac{ \sum \nolimits _{k=1}^{K}\left | c_{bk} \right |^{2} + z_{bk}^{c} }{K}.
\end{align}
\end{subequations}
with the other variables fixed. Finally, by alternately minimizing and repeatedly updating the upper-bound function {\eqref{equ_L_approx}}, we obtain the reweighted algorithm described in Algorithm {\ref{alg_jsbl_l2}}. As the objective function in \eqref{equ_L_approx} decreases or keeps unchanged in each iteration, the proposed JSBL-$\ell_2$ algorithm can promise a local minimum of problem~\eqref{equ_L_approx}.
\begin{algorithm}[!ht]
\label{alg_jsbl_l2}
\caption{JSBL-$\ell_2$ Algorithm}
\SetKwInOut{Input}{Input}
\SetKwInOut{Output}{Output}
\SetKwFor{While}{while}{}{}%
\Input{ $\boldsymbol{\Phi}$, $\mathbf{y}$, $\lambda$, $\beta$, $T_{\mathrm{JSBL}}$, $\epsilon$ .}
\While{$t\le T_{\mathrm{JSBL}}$ $\mathrm{or}$ $\left \| \mathbf{s}+\mathbf{c} - \mathbf{s}_{\mathrm {old}}-  \mathbf{c}_{\mathrm {old}}\right \|_{2}^{2} > \epsilon $}
{
1. $\mathbf{s}_{\mathrm {old}} = \mathbf{s}$ and $\mathbf{c}_{\mathrm {old}} = \mathbf{c}$. \\
2. Update $ \mathbf{s}^{\star }$ and $ \mathbf{c}^{\star }$ using (\ref{equ_opt_s_c}). \\
3. Update $ \left (\mathbf{z}^{s}  \right )^{\star }$ and $ \left (\mathbf{z}^{c}  \right )^{\star }$ using \eqref{equ_opt_zs_zc}. \\
4. Update $ \left (\gamma _{bk}^{s}  \right )^{\star }$ and $ \left (\gamma _{b}^{c}  \right )^{\star }$ using (\ref{equ_opt_gamma_s_c}). \\
5. $t = t + 1$.
}
\Output  {$\hat{\mathbf{x} } = \left ( \boldsymbol{\Gamma }^{s} + \mathbf{I}_{K} \otimes \boldsymbol{\Gamma }^{c}  \right )\boldsymbol{\Phi }^{H} \left ( \boldsymbol{\Sigma }^{sc} \right )^{-1} \mathbf{y}$.}
\end{algorithm}

\section{SIC-Based Hybrid Precoding through sub-array grouping}
As a result of CSI acquisition in Section IV, we now consider the channel is known at both TX and RX ends. The task of this section is to design the hybrid precoding and combining matrices for the DPA-MIMO system. The processed received signal after combining is given by
\begin{equation}
\label{equ_data_transmit}
{{\mathbf{y}}_{\mathrm{d}}} = {\mathbf{U}}_{{\mathrm{D}}}^H{\mathbf{U}}_{{\mathrm{R}}}^H{\mathbf{H}}{{\mathbf{V}}_{{\mathrm{R}}}}{{\mathbf{V}}_{{\mathrm{D}}}}{\mathbf{s}_{\mathrm{d}}} + {\mathbf{U}}_{{\mathrm{D}}}^H{\mathbf{U}}_{{\mathrm{R}}}^H{{\mathbf{z}}_{\mathrm{d}}} ,
\end{equation}
where $\mathbf{s}_{\mathrm d} \in \mathbb{C} ^{N_s \times 1}$ is the data vector such that $\mathbb{E} \left \{ \mathbf{s}_{\mathrm d} \mathbf{s}_{\mathrm d}^{H}  \right \} =\frac{{{P_{\text{d}}}}}{{{N_s}}}{{\mathbf{I}}_{{N_s}}}$, $P_{\mathrm d}$ is the average transmitting power, $\mathbf{z}_{\mathrm d}$ is a Gaussian noise vector with $ \mathcal{CN}\left ( \mathbf{0}, \sigma _{\mathrm z}^{2}\mathbf{I}_{N_r^{\mathrm {tot}}} \right )$, $\mathbf{V}_{\mathrm{R}}={\mathrm{blkdiag}}\left \{ {\mathbf{v}_{\mathrm{R}}^{\left [ 1 \right ]}},\cdots ,{\mathbf{v}_{\mathrm{R}}^{\left [ M_t \right ]}} \right \} $ and $\mathbf{U}_{\mathrm{R}}={\mathrm{blkdiag}}\left \{ {\mathbf{u}_{\mathrm{R}}^{\left [ 1 \right ]}},\cdots ,{\mathbf{u}_{\mathrm{R}}^{\left [ M_r \right ]}} \right\} $. Furthermore, we assume the amplitude of each element of ${\mathbf{v}_{\mathrm{R}}^{\left [ n \right ]}}$ and ${\mathbf{u}_{\mathrm{R}}^{\left [ m \right ]}}$ equals $\frac{1}{{\sqrt {N_t^{{\text{sub}}}} }}$ and $\frac{1}{{\sqrt {N_r^{{\text{sub}}}} }}$, respectively.  With Gaussian signaling employed at the TX, the instantaneous achievable SE is
\begin{equation}
\label{equ_rate_tx_rx}
\begin{aligned}
R = {\log _2}\Big| {{\mathbf{I}}_{{N_s}}} + \frac{{{P_{\text{d}}}}}{{{N_s}}} & {\mathbf{R}}_{\text{z}}^{ - 1}{\mathbf{U}}_{{\text{D}}}^H{\mathbf{U}}_{{\text{R}}}^H{\mathbf{H}}{{\mathbf{V}}_{{\text{R}}}}{{\mathbf{V}}_{{\text{D}}}}\\
 &\cdot {\mathbf{V}}_{{\text{D}}}^H{\mathbf{V}}_{{\text{R}}}^H{{\mathbf{H}}^H}{{\mathbf{U}}_{{\text{R}}}}{{\mathbf{U}}_{{\text{D}}}} \Big|,
\end{aligned}
\end{equation}
where ${{\mathbf{R}}_{\text{z}}} \triangleq \sigma _{\text{z}}^2{\mathbf{U}}_{{\text{D}}}^H{\mathbf{U}}_{{\text{R}}}^H{{\mathbf{U}}_{{\text{R}}}}{{\mathbf{U}}_{{\text{D}}}}$. Thus we determine the hybrid precoders and combiners by maximizing the SE defined in \eqref{equ_rate_tx_rx} with a transmitting power constraint $\left \| \mathbf{V}_{\mathrm{R}} \mathbf{V}_{\mathrm{D}} \right \|_{F}^{2} = N_s$. 
\subsection{Design of $\mathbf{V}_{\mathrm{D}}$ and $\mathbf{V}_{\mathrm{R}}$}
In lieu of maximizing the SE, we design $\mathbf{V}_{\mathrm{R}}$ and $\mathbf{V}_{\mathrm{D}}$ to maximize the mutual information achieved by Gaussian signaling~\cite{el2014spatially}. Then, the hybrid precoder design problem can be written as
\begin{subequations}
\label{problem_max_rate_tx}
\begin{alignat}{2}
& \underset{ \mathbf{V}_{\mathrm{R}},\mathbf{V}_{\mathrm{D}}}{\text{max}}
& & \log _{2} \left | \mathbf{I}_{N_r^{\mathrm{tot}}} + P_{\mathrm d}\left( N_s \sigma _{\mathrm z}^{2} \right)^{-1}  \mathbf{H} \mathbf{V}_{\mathrm{R}}\mathbf{V}_{\mathrm{D}} \mathbf{V}_{\mathrm{D}}^{H}\mathbf{V}_{\mathrm{R}}^H \mathbf{H}^{H} \right | \\
& \;\;\;\text{s.t.}
& \quad & \left \| \mathbf{V}_{\mathrm{R}} \mathbf{V}_{\mathrm{D}} \right \|_{F}^{2} = N_s\\
&&& | [{\mathbf{V}}_{\text{R}}  ]_{i,j}  | =  {1\mathord{\left/{\vphantom {1 2}} \right. \kern-\nulldelimiterspace} \sqrt{N_t^{\text{sub}}}},\; \forall i,j \in \mathcal{V}_{t}
\end{alignat}
\end{subequations}
where ${\mathcal{V}}_{t}$ denotes the set of non-zero element of ${\mathbf{V}}_{\text{R}} $.

\subsubsection{Design of $\mathbf{V}_{\mathrm{D}}$}
\begin{figure}[!htb]
\centering
\includegraphics[width=2.2in]{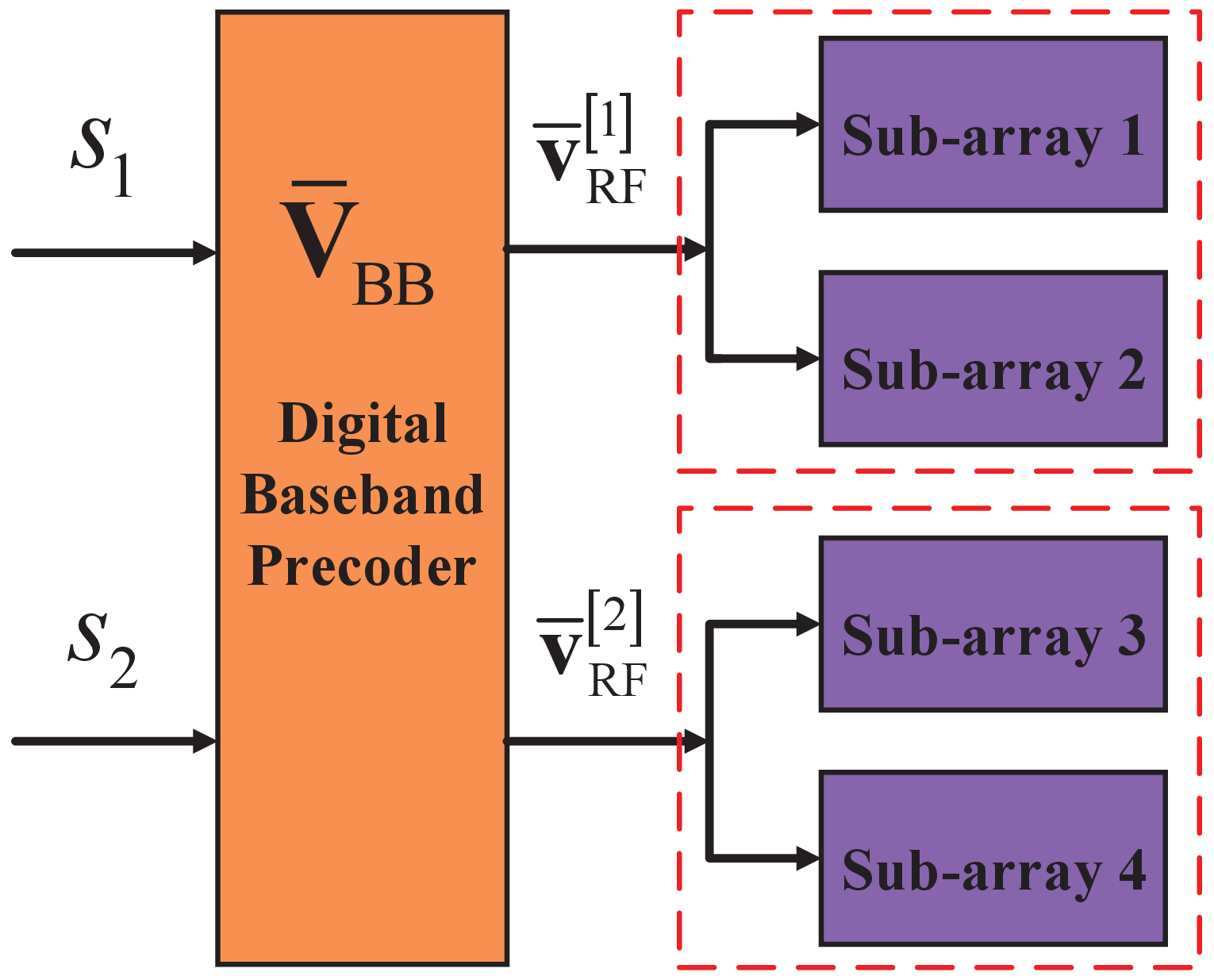}
\caption{Illustration of sub-array grouping with $N_s = 2$ and $M_t = 4$.}
\label{fig_subarray_grouping}
\end{figure}

We first group the sub-arrays at the TX according to the number of data streams $N_s$ satisfying $N_s \leq M_t$. We simply assign successive equal number of sub-arrays to each data stream if $M_t$ is multiples of $N_s$, otherwise, redundant sub-arrays are all assigned to any data stream. Note that this homogeneous grouping strategy is not a special case of the hybridly connected structure based partition strategy in~\cite{zhang2018hybridly}. We use the vector $\mathbf{d}_{t} \triangleq \left [ d_1, \cdots , d_{N_s} \right ]$ to indicate the number of sub-arrays assigned to each data stream. The optimal grouping strategy for the DPA-MIMO system is scheduled for future research. Furthermore, assisted by the combination of sub-array grouping and variable decoupling, the proposed scheme can handle both the TX and the RX design with a flexible number of data steams in comparison with~\cite{gao2016energy}.  Thus, we have the precoders with the new structures as $\bar{\mathbf{V}}_{\mathrm{D}} \in \mathbb{C}^{N_s\times N_s}$ and $\bar{\mathbf{V}}_{\mathrm{R}} = \mathrm{blkdiag} \left \{ \bar{\mathbf{v}}_{\mathrm{R}}^{\left [ 1 \right ]}, \cdots ,\bar{\mathbf{v}}_{\mathrm{R}}^{\left [ N_s \right ]} \right \} \in \mathbb{C}^{N_t^{\mathrm{tot}}\times N_s}$ where
\begin{equation}
\bar{\mathbf{v}}_{\mathrm{R}}^{\left [ i \right ]} =  \Big [ \Big ( \mathbf{v}_{\mathrm{R}}^{\left [ \sum_{j=1}^{i-1}d_j+1 \right ]} \Big )^T,\cdots , \Big ( \mathbf{v}_{\mathrm{R}}^{\left [ \sum_{j=1}^{i}d_j\right ]} \Big )^T \Big ]^T \in \mathbb{C}^{d_{i}N_t^{\mathrm{sub}}\times 1}.
\end{equation}
In Fig.{~\ref{fig_subarray_grouping}}, we give a example of sub-array grouping with $N_s = 2$ and $M_t = 4$. In this scenario, two data symbols are first processed by a $2 \times 2$ digital precoding matrix. Then, each precoded data is allocated to two sub-arrays. Such a grouping operation facilitates the efficient design of RF precoding using SIC, as will be shown in the following.

With the defined variables $\bar{\mathbf{V}}_{\mathrm{R}}$ and $\bar{\mathbf{V}}_{\mathrm{D}}$ substituted into \eqref{problem_max_rate_tx}, we follow the common decoupling procedure~\cite{park2017dynamic} that given a fixed RF precoder $\bar{\mathbf{V}}_{\mathrm{R}}$ and an equivalent channel matrix $\mathbf{H}_{\mathrm{eq}} \triangleq \mathbf{H} \bar{\mathbf{V}}_{\mathrm{R}}$, the optimal digital precoder has a closed-form water-filling solution as
\begin{equation}
\label{equ_waterfilling}
\bar{\mathbf{V}}_{\mathrm{D}} ^{\star } = \mathbf{D}_t^{-1/2} \mathbf{U}_{\mathrm{e}} \boldsymbol{\Lambda }_{\mathrm{e}} ^{1/2}  ,
\end{equation}
where $\mathbf{D}_t = \bar{\mathbf{V}}_{\mathrm{R}}^{H}\bar{\mathbf{V}}_{\mathrm{R}} = \mathrm{diag}\left \{ \mathbf{d}_{t} \right \}$, $\mathbf{U}_{\mathrm{e}}$ is the set of right singular vectors corresponding to the $N_s$ largest singular values of  $\mathbf{H}_{\mathrm {eq}}\mathbf{D}_t^{-1/2} $ and $\boldsymbol{\Lambda }_{\mathrm{e}}$ is a diagonal matrix with the allocated powers to each data stream on its main diagonal. 

\subsubsection{Design of $\mathbf{V}_{\mathrm{R}}$}
With an equal power allocation scheme, i.e., $\boldsymbol{\Lambda }_{\mathrm{e}} \approx \mathbf{I}_{N_s}$, which shows a little loss in performance for moderate and high SNR regimes, we obtain the approximately optimal digital precoder $\bar{\mathbf{V}}_{\mathrm{D}}  \approx  \mathbf{D}_t^{-1/2} \mathbf{U}_{\mathrm{e}}$. Thus, the RF precoder can be obtained by solving the following problem
\begin{subequations}
\label{problem_FRF_design}
\begin{alignat}{2}
& \underset{ \tilde{\mathbf{V}}_{\text{R}} }{\text{max}}
& &  \log _{2} \left | \mathbf{I}_{N_r^{\mathrm{tot}}} + P_{\mathrm d}\left( N_s \sigma _{\mathrm z}^{2} \right)^{-1}  \mathbf{H} \tilde{\mathbf{V}}_{\mathrm{R}} \tilde{\mathbf{V}}_{\mathrm{R}}^{H} \mathbf{H}^{H} \right |  
\label{problem_FRF_design_objective}\\
&\; \, \text{s.t.}
& \quad &  | [\tilde{\mathbf{V}}_{\text{R}}  ]_{i,j}  | =  {1\mathord{\left/{\vphantom {1 2}} \right. \kern-\nulldelimiterspace} \sqrt{d_j N_t^{\text{sub}}}},\; \forall i,j \in \tilde{\mathcal{V}}_{t}
\end{alignat}
\end{subequations}
where $\tilde{\mathbf{V}}_{\mathrm{R}} \triangleq \bar{\mathbf{V}}_{\mathrm{R}}  \mathbf{D}_t^{-1/2} \in \mathbb{C}^{N_t^{\mathrm{tot}}\times N_s}$ and $ \tilde{\mathcal{V}}_{t}$ denotes the set of non-zero element of $\tilde{\mathbf{V}}_{\text{R}} $.

According to the block structure of $\tilde{\mathbf{V}}_{\mathrm{R}} $, it is observed that problem {\eqref{problem_FRF_design}} with nonconvex constraints can be decomposed into a series of simple subproblems, each of which only considers one specific group of sub-arrays~\cite{gao2016energy,zhang2018hybridly}. In particular, we can divide the matrix $\tilde{\mathbf{V}}_{\mathrm{R}}$ as $\tilde{\mathbf{V}}_{\mathrm{R}} = \left [ \tilde{\mathbf{V}}_{N_s -1} \tilde{\mathbf{v}}_{N_s}\right ]$, where $\tilde{\mathbf{v}}_{N_s}$ is the $N_s$th column and $\tilde{\mathbf{V}}_{N_s -1}$ is a matrix containing the first $N_s -1$ columns of $\tilde{\mathbf{V}}_{\mathrm{R}}$, respectively. Thus, the cost function of {\eqref{problem_FRF_design}} can be written as
\begin{equation}
\label{equ_R_decomp}
\eqref{problem_FRF_design_objective} 
= \sum\nolimits_{n=1}^{N_s} \log _{2} \left | 1 + P_{\mathrm d}\left( N_s \sigma _{\mathrm z}^{2} \right)^{-1} \tilde{\mathbf{v}}_{n}^{H} \mathbf{H}^{H} \mathbf{R}_{n -1}^{-1} \mathbf{H}   \tilde{\mathbf{v}}_{n}    \right |,
\end{equation}
where $\mathbf{R}_{n -1} = \mathbf{I}_{N_r^{\mathrm{tot}}} + P_{\mathrm d}\left( N_s \sigma _{\mathrm z}^{2} \right)^{-1} \mathbf{H}  \tilde{\mathbf{V}}_{n -1} \tilde{\mathbf{V}}_{n -1}^{H} \mathbf{H}^{H}$ and $\mathbf{R}_{0} = \mathbf{I}_{N_r^{\mathrm{tot}}}$. This equality follows from $\left | \mathbf{I} + \mathbf{A} \mathbf{B} \right | = \left | \mathbf{I} + \mathbf{B} \mathbf{A} \right | $ and $\left |\mathbf{A} \mathbf{B}  \right | =\left |\mathbf{A}  \right | \left |\mathbf{B}  \right |$~\cite{kaare2000cookbook}.

In \eqref{equ_R_decomp}, the total achievable rate can be a summation of the subrates of all data streams. Motivated by the idea of SIC for multi-user detection, we can first optimize the achievable subrate of the first data stream and then update $\mathbf{R}_1$. After some iterations, the optimal RF precoder design subproblem for the $n$th group of sub-arrays can be equivalently represented by
\begin{equation}
\label{problem_opt_analog}
\tilde{\mathbf{v}}_{n}^{\star} = \text{arg} \; \underset{\tilde{\mathbf{v}}_{n}}{\text{max}} \log _{2} \left | 1 + P_{\mathrm d}\left( N_s \sigma _{\mathrm z}^{2} \right)^{-1} \tilde{\mathbf{v}}_{n}^{H} \mathbf{T}_{n-1} \tilde{\mathbf{v}}_{n}    \right |,
\end{equation}
where $\mathbf{T}_{n-1} = \mathbf{H}^{H} \mathbf{R}_{n -1}^{-1} \mathbf{H}$. Note that $\mathbf{T}_{n}$ can be iteratively obtained without matrix inverse, with its derivation shown in Appendix~\ref{appendix_derivation_2}. Due to the special structure of $\tilde{\mathbf{v}}_{n}$, (\ref{problem_opt_analog}) can be further simplified to
\begin{equation}
\label{problem_opt_analog_structure}
\left( \bar{\mathbf{v}}_{\mathrm{R}}^{\left [ n \right ]} \right)^{\star} = \text{arg} \; \underset{\bar{\mathbf{v}}_{\mathrm{R}}^{\left [ n \right ]}}{\text{max}} \; \left( \bar{\mathbf{v}}_{\mathrm{R}}^{\left [ n \right ]} \right)^{H} \hat{\mathbf{T}}_{n-1} \bar{\mathbf{v}}_{\mathrm{R}}^{\left [ n \right ]}    ,
\end{equation}
where $\hat{\mathbf{T}}_{n-1}$ is a $d_n N_t^{\mathrm{sub}} \times d_n N_t^{\mathrm{sub}}$ Hermitian matrix formed as a submatrix of matrix $\mathbf{T}_{n-1}$ by taking the $\left ( N_t^{\mathrm{sub}} \sum_{i=1}^{n-1}d_i+1 \right)$th row and column to the $\left ( N_t^{\mathrm{sub}} \sum_{i=1}^{n}d_i \right)$th row and column of $\mathbf{T}_{n-1}$. Since each element of $\bar{\mathbf{v}}_{\mathrm{R}}^{\left [ n \right ]}$ can be separated in \eqref{problem_opt_analog_structure}, its optimal solution can be obtained by iteratively updating the following equation until convergence
\begin{equation}
\label{eq_opt_f_n}
{\left[ \bar{\mathbf{v}}_{\mathrm{R}}^{\left [ n \right ]}  \right]_i} = \frac{1}{{\sqrt {N_t^{{\text{sub}}}} }} e^{j \measuredangle \left(  {\sum\limits_{j \ne i} {{{\left[ {{{\hat {\mathbf{T}}}_{n - 1}}} \right]}_{i,j}}{{\left[\bar{\mathbf{v}}_{\mathrm{R}}^{\left [ n \right ]}  \right]}_j}} } \right) } ,
\end{equation}
where the symbol $\measuredangle\left(. \right) $ extracts the corresponding phases of the element. Note that this iterative procedure is guaranteed to converge to a local optimum since the objective function of \eqref{problem_opt_analog_structure} increases in each iteration.

\subsection{Design of $\mathbf{U}_{\mathrm{D}}$ and $\mathbf{U}_{\mathrm{R}}$}
\subsubsection{Design of $\mathbf{U}_{\mathrm{R}}$}
We group the sub-arrays at the RX in the same way as the TX. Furthermore, we decouple the design of $\bar{\mathbf{U}}_{\mathrm{R}}$ and $\bar{\mathbf{U}}_{\mathrm{D}}$ by first optimizing the RF combiner with assumed ideal digital combiner and then finding the optimal digital combiner for the obtained RF combiner. 
Since $\bar{\mathbf{U}}_{\mathrm{R}}^{H}\bar{\mathbf{U}}_{\mathrm{R}} = \mathbf{D}_r = \mathrm{diag} \left\{ \mathbf{d}_r \right\}$ where $\mathbf{d}_r \triangleq \left [ \bar{d}_1,\cdots ,\bar{d}_{N_s} \right ]$, we can define $\tilde{\mathbf{U}}_{\mathrm{R}} \triangleq \bar{\mathbf{U}}_{\mathrm{R}}{\mathbf{D}}_r^{-1/2}$, which results in the similar problem as (\ref{problem_FRF_design}):
\begin{subequations}
\label{problem_WRF_design}
\begin{alignat}{2}
& \underset{ \tilde{\mathbf{U}}_{\mathrm{R}}}{\text{max}}
& &  \log _{2} \left | \mathbf{I}_{N_s} + P_{\mathrm d}\left( N_s \sigma _{\mathrm z}^{2} \right)^{-1}  \tilde{\mathbf{U}}_{\mathrm{R}}^{H}\bar{\mathbf{H}}_{\mathrm {eq}}   \bar{\mathbf{H}}_{\mathrm {eq}}^{H}   \tilde{\mathbf{U}}_{\mathrm{R}} \right | \\
&\;  \, \text{s.t.}
& \quad &  | [\tilde{\mathbf{U}}_{\text{R}}  ]_{i,j}  | =  {1\mathord{\left/{\vphantom {1 2}} \right. \kern-\nulldelimiterspace} \sqrt{\bar{d}_j N_r^{\text{sub}}}},\; \forall i,j \in \mathcal{V}_{r}
\end{alignat}
\end{subequations}
where $ \mathcal{V}_{r}$ denotes the set of non-zero element of $\tilde{\mathbf{U}}_{\text{R}} $.

\subsubsection{Design of $\mathbf{U}_{\mathrm{D}}$}
Assuming all other beamformers are fixed, the optimal digital combiner based on the minimum mean-square error (MMSE) criterion  is formulated as~\cite{el2014spatially}
\begin{equation}
\label{equ_WBB_opt}
\bar{\mathbf{U}}_{\mathrm{D}}^{\star} 
 = \left (P_{\mathrm{d}}/N_s  \right ) \mathbf{J}^{-1} \bar{\mathbf{U}}_{\mathrm{R}}^H \mathbf{H} \bar{\mathbf{V}}_t,
\end{equation}
where $\mathbf{J} = \frac{P_{\mathrm{d}}}{N_s} \bar{\mathbf{U}}_{\mathrm{R}}^H \mathbf{H} \bar{\mathbf{V}}_t  \bar{\mathbf{V}}_t^{H}   \mathbf{H}^H \bar{\mathbf{U}}_{\mathrm{R}}  + \sigma _{\mathrm{z}}^2 \mathbf{D}_r  \in \mathbb{C}^{N_s \times N_s}$ and $\bar{\mathbf{V}}_t = \bar{\mathbf{V}}_{\mathrm{R}}^{\star} \bar{\mathbf{V}}_{\mathrm{D}}^{\star}$. Finally, the proposed hybrid precoding approach is summarized in Algorithm~{\ref{alg_hybrid_precoding}}.
\begin{algorithm}[!ht]
\label{alg_hybrid_precoding}
\caption{SIC-Based Hybrid Precoding Through Sub-array Grouping}
\SetKwInOut{Input}{Input}
\SetKwInOut{Output}{Output}
\SetKwFor{While}{while}{}{}%
\Input{ ${N_s}$, $P_{\text{d}}$ and $\sigma _{\text{z}}^2$.}
1. Group the sub-arrays at the TX and RX respectively. \\
2. Update ${\hat {\mathbf{v}}_n}$ using \eqref{eq_opt_f_n} until convergence . \\
3. Optimize $\bar{\mathbf{V}}_{\mathrm{D}} ^{\star }$ using \eqref{equ_waterfilling}. \\
4. Optimize $\tilde{\mathbf{U}}_{\mathrm{R}}^{\star }$  the same way as \eqref{problem_FRF_design}. \\
5. Optimize $\bar{\mathbf{U}}_{\mathrm{D}} ^{\star }$ using \eqref{equ_WBB_opt}. \\
\Output  {$\mathbf{V}_{\mathrm{R}} ^{\star }$, $\mathbf{V}_{\mathrm{D}} ^{\star }$, $\mathbf{U}_{\mathrm{R}} ^{\star }$ and $\mathbf{U}_{\mathrm{D}} ^{\star }$.}
\end{algorithm}
\section{Simulation Results}
The performance of the proposed algorithms is evaluated through simulation with the following parameters. The TX and RX sub-arrays are ULAs of half-wavelength antenna spacing. For the sub-array spacing, we simply set $d_a = 9\lambda_c $ for algorithm verification. Note that the practical value of $d_a$ has relationship with joint sparsity and it is worth pointing out that this relationship needs to be validated through extensive measurement and study of mmWave channels~\cite{huo20175g,huo2018cellular}. Since the amplitude of the LoS components is typically 5 to $10\;\text{dB}$ stronger than that of the NLoS components at mmWave frequencies~\cite{fan2018angle}, the channel coefficients are generated through \eqref{equ_channel_subarray} with the variances of the channel paths as $\sigma_{\mathrm{LoS}}^{2} = 1 $ and $\sigma_{\mathrm{NLoS}}^{2} = 10^{-0.5}$~\cite{gao2017reliable}. We denote the channel common sparsity as $L_c = \left | \Omega _c \right |\geq 1$ and assume the equal channel individual sparsity among different sub-arrays as $L = \left | \Omega _{m,n} \right | + \left | \Omega _c \right | = 5$ for simulation convenience.  For the generation of the  proposed DPA-MIMO channel, we first generate $L_c$ common paths one of which is the LoS path for all the sub-arrays. For any common path, the phase variation across the sub-arrays at the TX/RX induced by the sub-array spacing should be considered~\cite{singh2015feasibility}, i.e., for common path $i$ with $\vartheta _{m,n}^{\left( i \right)} = {\vartheta ^{\left( i \right)}}$ and $\psi _{m,n}^{\left( i \right)} = {\psi ^{\left( i \right)}}$, its complex amplitude is calculated by
$\alpha _{m,n}^{\left( i \right)} = \alpha _{1,1}^{\left( i \right)}{e^{ - j\left( {\varphi _{r,m}^{\left( i \right)} + \varphi _{t,n}^{\left( i \right)}} \right)}}$ where $\varphi _{r,m}^{\left( i \right)} = \frac{{2\pi }}{{{\lambda _c}}}\left( {m - 1} \right)\left( {{d_a} + \left( {N_r^{{\text{sub}}} - 1} \right){d_e}} \right)\cos \left( {{\vartheta ^{\left( i \right)}}} \right)$ and $\varphi _{t,n}^{\left( i \right)} = \frac{{2\pi }}{{{\lambda _c}}}\left( {n - 1} \right)\left( {{d_a} + \left( {N_t^{{\text{sub}}} - 1} \right){d_e}} \right)\cos \left( {{\psi ^{\left( i \right)}}} \right)$.
Then, we generate independent $\left ( L-L_c \right )$ paths for each pair of the TX and RX sub-arrays.

In Algorithm {\ref{alg_jomp}}, the threshold parameters are set to be $\delta _1 = N \sigma _{\mathrm{z}}^2$ and $\delta _2 = 0.1 N \sigma _{\mathrm{z}}^2$; the maximal iteration numbers are chosen as $T_{\mathrm{1,JOMP}} = L_c$ and $T_{\mathrm{2,JOMP}} = \left (L+2-T_{\mathrm{1,JOMP}}  \right )K$. In Algorithm \ref{alg_jsbl_l2}, we set the weight $\beta = 3.3$, the iteration number $T_{\mathrm{JSBL}} = 80$ and the error tolerance $\epsilon = 10^{-4}$. In the following, two types of SNRs are considered: one is the pilot-to-noise ratio (PNR) defined as $10\log_{10} \left( P_{\mathrm{p}} / \sigma_{\mathrm{z}}^2 \right)$, and the other is the data-to-noise ratio (DNR) defined as $10\log_{10} \left( P_{\mathrm{d}} / \sigma_{\mathrm{z}}^2 \right)$. The performance metric for channel estimation is the normalized MSE (NMSE) defined as $\label{equ_nmse}
{\text{NMSE}} \triangleq 10{\log _{10}}\Big( {\mathbb{E}\Big\{ {\frac{1}{{N_t^{{\text{tot}}}N_r^{{\text{tot}}}}}\Big\| {{\mathbf{H}} - \hat {\mathbf{H}}} \Big\|_F^2} \Big\}} \Big).$ The hybrid precoding schemes based on the channel estimates $\hat {\mathbf{H}}$ are evaluated through the SE defined in {\eqref{equ_rate_tx_rx}.

\begin{table*}[!ht]
\small
\caption{Comparision of Pilot Overhead and Complexity for Channel Estimation Algorithms} 
\centering 
\begin{threeparttable}[b]
\begin{tabular}{c c c c}
\toprule[1.5pt]
\textbf{Category} &  \textbf{Algorithm} & \textbf{Pilot Beam Overhead} &   \textbf{Computational Complexity}\tnote{*} \\
\midrule
\textbf{Non-CS} &\textbf{DFTB} & $N_t^{{\text{tot}}} N_r^{{\text{tot}}}$ & $\mathcal{O}\left( {T_{{\text{DFTB}}}}K{B^2} \right)$ \\
\midrule
\multirow{4}{*}{\textbf{CS}}
& \textbf{OMP} & \multirow{4}{*}{$N_t^{{\text{beam}}} N_r^{{\text{beam}}}$} &  $\mathcal{O}\left( T_{{\text{OMP}}}^3N \right)$ \\
&  \textbf{JOMP} (proposed) & & $\mathcal{O}\left( (T_{{\text{1,OMP}}}^3 + T_{{\text{1,OMP}}}^3)N \right)$ \\
& \textbf{SBL} & &  $\mathcal{O}\left( {T_{{\text{SBL}}}}{N^3} \right)$\\
&\textbf{JSBL-$\ell_2$} (proposed) & &  $\mathcal{O}\left( {T_{{\text{JSBL}}}}{N^3} \right)$ \\
\bottomrule[1.5pt]
\end{tabular}
\label{table_complexity}
\begin{tablenotes}
\item [*] The related symbols are listed as : $ N \triangleq N_t^{\mathrm{beam}} N_r^{\mathrm{beam}}$, $ B \triangleq N_t^{\mathrm{sub}} N_r^{\mathrm{sub}}$ and $K \triangleq M_t M_r$. Besides, $T_{{\text{DFTB}}}$, $T_{{\text{OMP}}}$, $T_{{\text{1,JOMP}}} \left( T_{{\text{2,JOMP}}}\right)$, $T_{{\text{SBL}}}$ and $T_{{\text{JSBL}}}$ are the iteration numbers for DFTB, OMP, JOMP, SBL, JSBL-$\ell_2$, respectively.
\end{tablenotes}
\end{threeparttable}
\end{table*}

This section consists of the following two parts: 
\begin{itemize}
\item In the first part, we compare the NMSE of the proposed JOMP and JSBL-$\ell_2$ estimators with the conventional OMP and SBL estimators by employing the designed training beam patterns. Additionally, the non-cooperative DFT codebook based channel estimator, named as DFTB, is served as a benchmark. The comparison of pilot beam overhead and computational complexity for the channel estimators is summarized in Table~\ref{table_complexity}. On the one hand, the SBL and the proposed JSBL-$\ell _2$ algorithms require higher complexity than the DFTB, the OMP and the proposed JOMP algorithms. On the other hand, the CS-based algorithms embrace less pilot beam overhead than the DFTB algorithm when $N_t^{{\text{beam}}}N_r^{{\text{beam}}} < N_t^{{\text{tot}}}N_r^{{\text{tot}}}$. 
\item In the second part, we investigate the performance of the proposed hybrid precoding scheme termed as GSIC, the SDR based alternating optimization (SDR-AO) scheme~\cite{yu2016alternating}, the hybrid precoding scheme~\cite{park2017dynamic} and the optimal fully-digital (FD) precoding scheme. We further compare the SE realized through the proposed hybrid beamformers based on the channel estimates obtained in the previous part.
\end{itemize}

\subsection{Performance Evaluation of Channel Estimation}
\begin{figure}[!htb]
\centering
\includegraphics[width=3.3in]{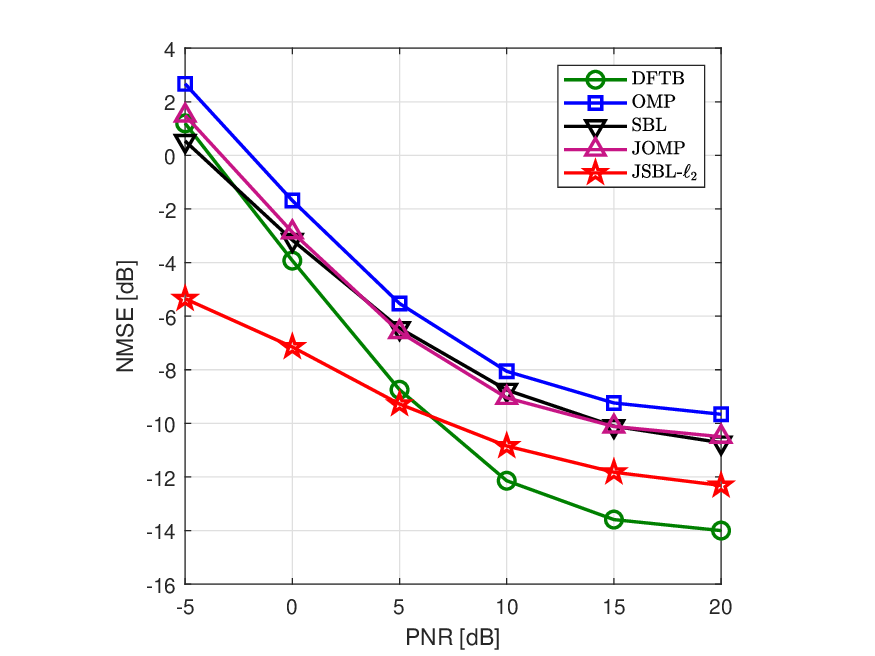}
\caption{NMSE versus PNR when $M_t = M_r = 4$, $N_t^{\text{sub}} = N_r^{\text{sub}} = 12$, $L = 5$, $L_c = 3$, $N_t^{\text{beam}} = 24$ and $N_r^{\text{beam}} = 36$.}
\label{fig_nmse_vs_snr}
\end{figure}
In Fig.{~\ref{fig_nmse_vs_snr}}, we compare the NMSE versus PNR for a DPA-MIMO setup with $M_t = M_r = 4$, $N_t^{\mathrm{sub}} = N_r^{\mathrm{sub}} = 12$, $N_t^{\mathrm{beam}} = 24$, $N_r^{\mathrm{beam}} = 36$ (partial-training case), $L_c = 3$ and $L = 5$. It is notable that the JSBL-$\ell_2$ estimator substantially outperforms the other estimators at low PNR. We further observe that the greedy JOMP estimator almost achieves the same channel estimation performance as the SBL estimator. At high PNR, the DFTB estimator with the full-training overhead achieves better performance over the other CS-based methods.

\begin{figure}[!htb]
\centering
\includegraphics[width=3.3in]{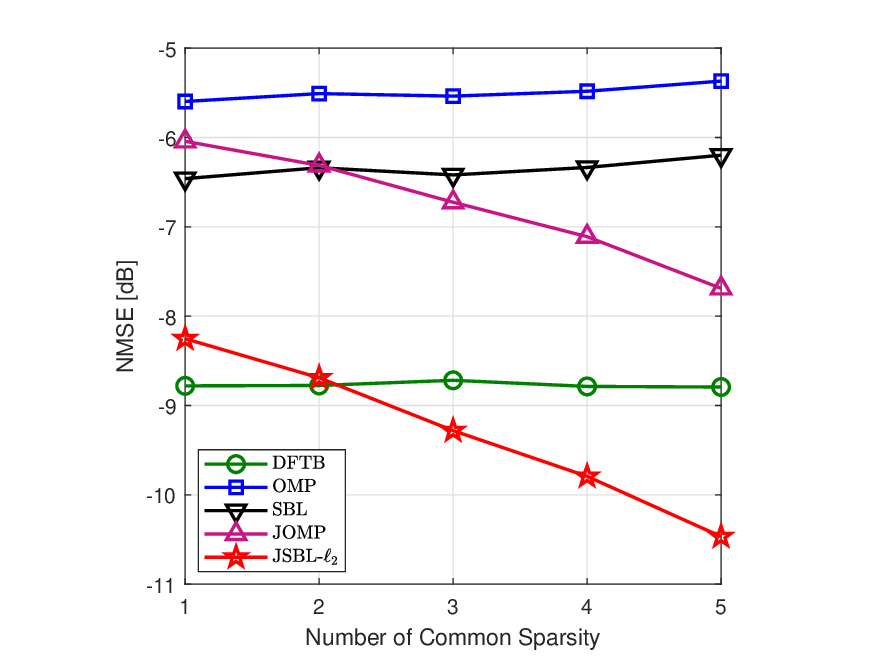}
\caption{NMSE versus common sparsity $L_c$ when $M_t = M_r = 4$, $N_t^{\text{sub}} = N_r^{\text{sub}} = 12$, $L = 5$, $\text{PNR} = 5 \;\text{dB}$, $N_t^{\text{beam}} = 24$ and $N_r^{\text{beam}} = 36$.}
\label{fig_nmse_vs_supp}
\end{figure}
In Fig.{~\ref{fig_nmse_vs_supp}}, we show the NMSE versus the common sparsity $L_c$ varying from 1 to $L$ for a DPA-MIMO setup with $M_t = M_r = 4$, $N_t^{\mathrm{sub}} = N_r^{\mathrm{sub}} = 12$, $N_t^{\mathrm{beam}} = 24$, $N_r^{\mathrm{beam}} = 36$ (partial-training case) and ${\text{PNR}} = \;5\;{\text{dB}}$. For both the proposed JOMP and JSBL-$\ell_2$ estimators, better channel estimation performance is obtained with an increasing number of the common support $L_c$, while the OMP, SBL and DFTB estimators keep the constant NMSE. This is because the two customized estimators take advantage of the jointly sparse characteristic of the DPA-MIMO channel to enhance the quality of estimated channels. Moreover, for $L_c \geq 3$, the JSBL-$\ell_2$ estimator surpasses the DFTB estimator in the accuracy of channel estimates.

\begin{figure}[!htb]
\centering
\subfloat[$\text{PNR} = 5 \;\text{dB}$]{\includegraphics[width=3.3in]{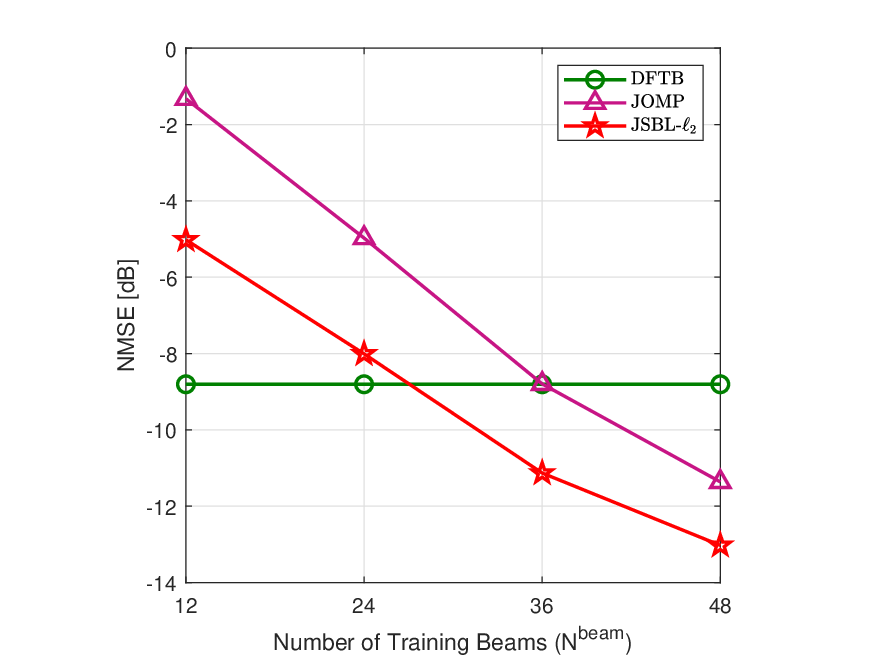}
\label{fig_nmse_vs_beam_5dB}} \\
\subfloat[$\text{PNR} = 20 \;\text{dB}$]{\includegraphics[width=3.3in]{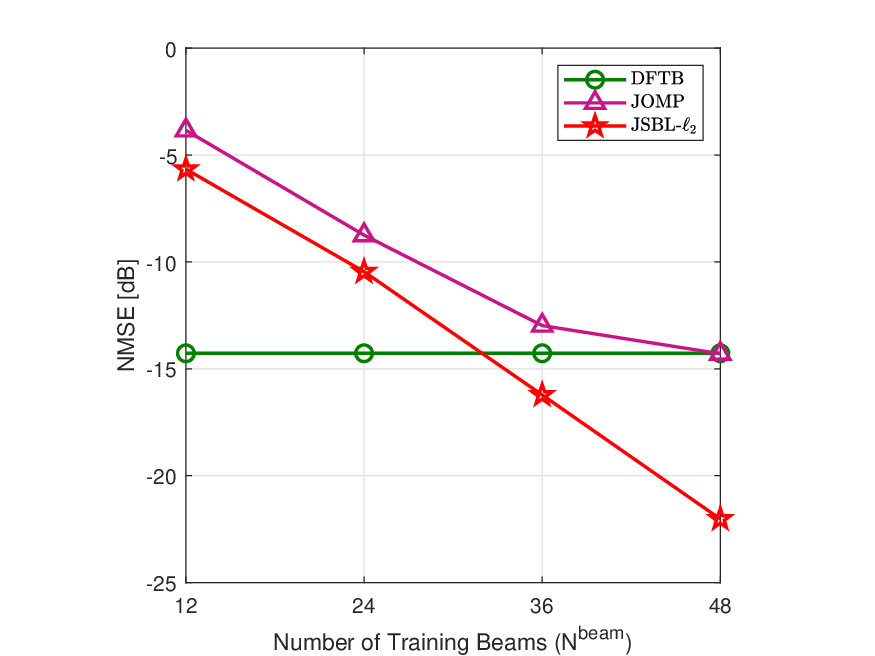}
\label{fig_nmse_vs_beam_20dB}}
\caption{NMSE versus number of training beams $N^{\text{beam}} = N_t^{\text{beam}} = N_r^{\text{beam}}$ when $M_t = M_r = 4$, $N_t^{\text{sub}} = N_r^{\text{sub}} = 12$, $L = 5$ and $L_c = 3$.}
\label{fig_nmse_vs_beam}
\end{figure} 
In Fig.{~\ref{fig_nmse_vs_beam}}, we further investigate that with how much training the proposed two CS-based methods can be competitive to the full-training based DFTB estimator for different PNRs. The number of training beams at both ends is assumed to the same as $N^{\mathrm{beam}}$ for a DPA-MIMO setup with $M_t = M_r = 4$, $N_t^{\mathrm{sub}} = N_r^{\mathrm{sub}} = 12$, $L_c = 3$ and $L = 5$. In Fig.{~\subref*{fig_nmse_vs_beam_5dB}},  with an increasing $N^{\mathrm{beam}}$, the NMSE of the proposed two estimators decreases monotonically at ${\text{PNR}} = \;5\;{\text{dB}}$. More specifically, the JOMP and JSBL-$\ell_2$ estimators can realize the nearly same reconstruction accuracy of the DFTB estimator with its $\frac{3}{4} \times \frac{3}{4} = 56.25\% $ and $\frac{1}{2} \times \frac{1}{2} = 25\% $ training overhead, respectively.  As shown in Fig.{~\subref*{fig_nmse_vs_beam_20dB}} at ${\text{PNR}} = \;20\;{\text{dB}}$, the JOMP estimator finally  approaches the same NMSE value as the DFTB estimator with $N^{\mathrm{beam}} = 48$, while the JSBL-$\ell_2$ estimator still maintains the performance advantage with a high PNR. 

\subsection{Performance Evaluation of Hybrid Precoding}
\begin{figure}[!htb]
\centering
\subfloat[$M_t = M_r = 3$]{\includegraphics[width=3.3in]{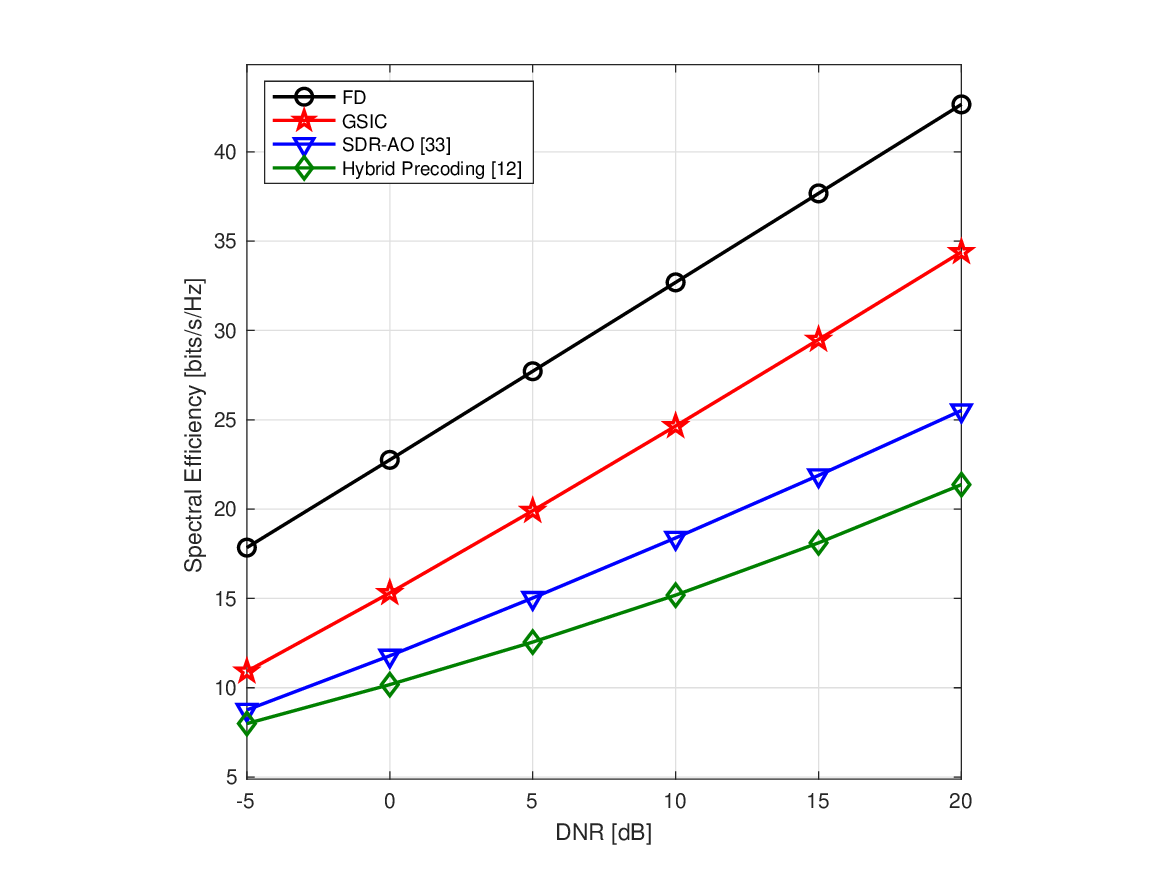}
\label{fig_hybrid_precoding_perfect_csi_M_3_Ns_3}} \\
\subfloat[$M_t = M_r = 6$]{\includegraphics[width=3.3in]{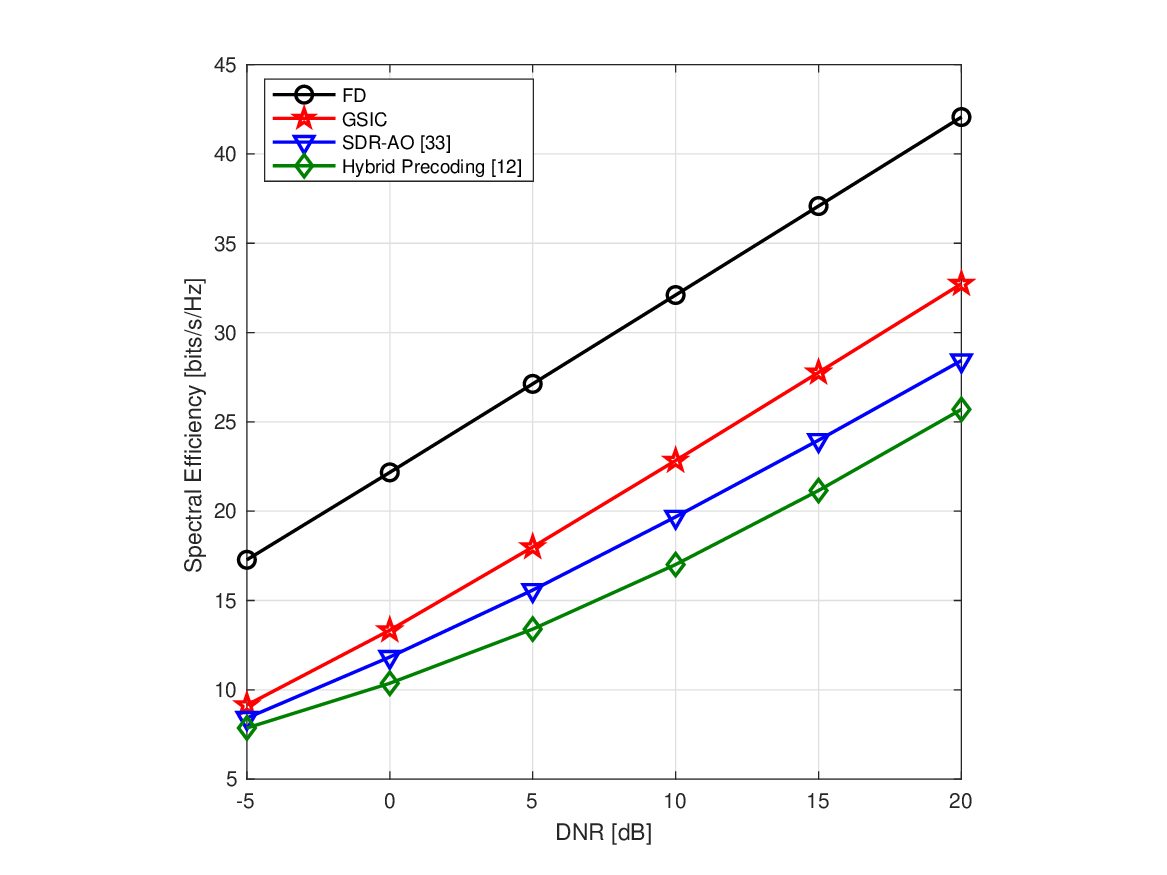}
\label{fig_hybrid_precoding_perfect_csi_M_6_Ns_3}}
\caption{SE versus DNR with perfect CSI when $N_t^{\text{tot}} = N_r^{\text{tot}} = 90$ and $N_s = 3$.}
\end{figure}
First, we compare the SE versus DNR for a various number of sub-arrays at the TX/RX with perfect CSI and a fixed number of total antennas. We set  the parameters as $N_t^{\mathrm{tot}} = N_r^{\mathrm{tot}} = 90$, $L_c = 3$ and $L = 5$. It is observed from Fig.{~\subref*{fig_hybrid_precoding_perfect_csi_M_3_Ns_3}} and Fig.{~\subref*{fig_hybrid_precoding_perfect_csi_M_6_Ns_3}} that there are distinct gaps between the FD scheme and the other three hybrid precoding schemes, which results from the beamforming gain loss of the sub-array based structure compared with the FD structure~\cite{yu2016alternating}. We further find that the proposed GSIC scheme can successfully surpass the SDR-AO and SVD-based schemes. The efficiency of the proposed GSIC scheme is verified through average central processing unit processing time. By using the same configuration as the Matlab experiment demonstrated in Fig.{~\subref*{fig_hybrid_precoding_perfect_csi_M_6_Ns_3}}, the proposed GSIC scheme almost run 115 times faster than the SDR-AO scheme~\cite{yu2016alternating} while only about 1.5 times slower than the hybrid precoding scheme in~\cite{park2017dynamic}. Therefore, in the following, we only employ the proposed GSIC scheme to design the hybrid beamformers based on the estimated channels obtained in Subsection VI-A.

\begin{figure}[!htb]
\centering
\includegraphics[width=3.3in]{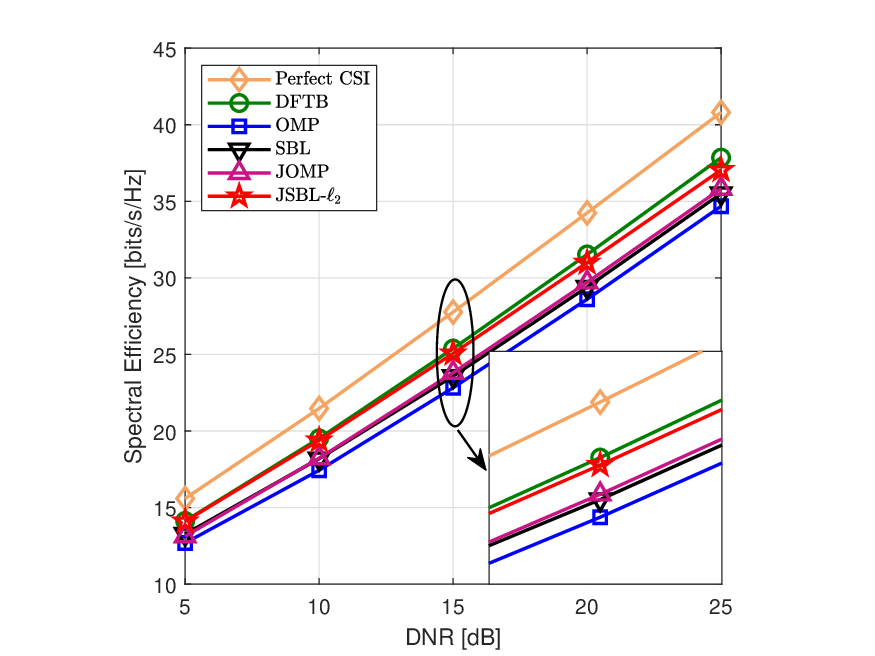}
\caption{SE versus DNR when $M_t = M_r = 4$, $N_t^{\text{sub}} = N_r^{\text{sub}} = 12$, $N_s = 4$, $L = 5$, $L_c = 3$, $\text{PNR} = 5 \;\text{dB}$, $N_t^{\text{beam}} = 24$ and $N_r^{\text{beam}} = 36$.}
\label{fig_rate_csi_vs_snr}
\end{figure}
In Fig.{~\ref{fig_rate_csi_vs_snr}}, we investigate the SE versus DNR for a DPA-MIMO setup with $M_t = M_r = 4$, $N_t^{\text{sub}} = N_r^{\text{sub}} = 12$, $N_s = 4$, $L = 5$, $L_c = 3$, $\text{PNR} = 5 \;\text{dB}$, $N_t^{\text{beam}} = 24$ and $N_r^{\text{beam}} = 36$. 
Obviously,  the improved channel estimation accuracy of the proposed algorithms
are reflected in their enhanced SEs.
Furthermore, the SE curve of the proposed JSBL-$\ell_2$ estimator nearly coincides with that of the DFTB estimator. In addition, the proposed less-complexity JOMP estimator can provide the same SE as the SBL estimator. 

\begin{figure}[!htb]
\centering
\includegraphics[width=3.3in]{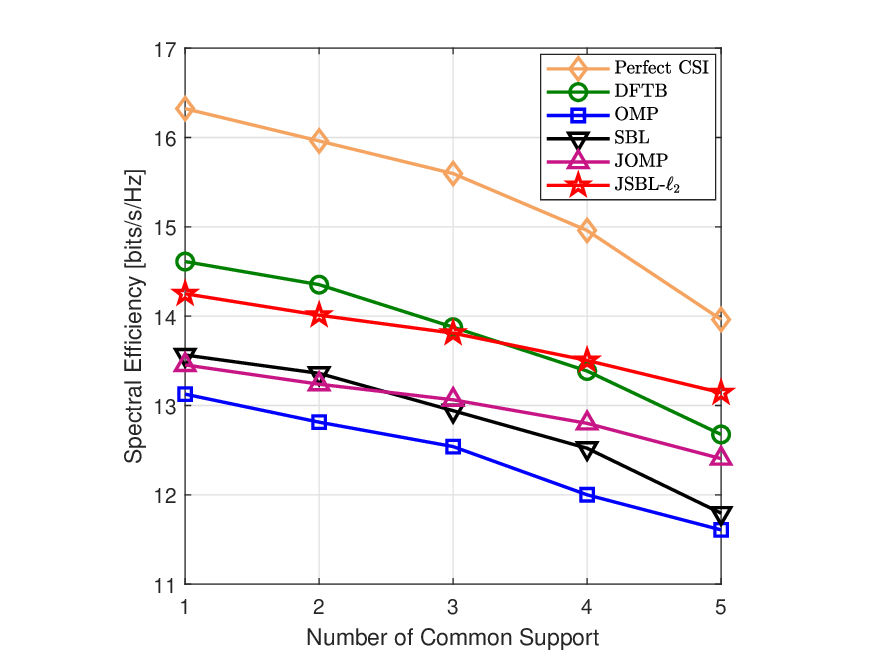}
\caption{SE versus number of common sparsity when $M_t = M_r = 4$, $N_t^{\text{sub}} = N_r^{\text{sub}} = 12$, $N_s = 4$, $L = 5$, $\text{DNR} = 5 \;\text{dB}$, $\text{PNR} = 5 \;\text{dB}$, $N_t^{\text{beam}} = 24$ and $N_r^{\text{beam}} = 36$.}
\label{fig_rate_csi_vs_supp}
\end{figure}
In Fig.{~\ref{fig_rate_csi_vs_supp}}, we present the SE versus the common sparsity $L_c$ varing from 1 to $L$ for a DPA-MIMO setup with $M_t = M_r = 4$, $N_t^{\text{sub}} = N_r^{\text{sub}} = 12$, $N_s = 4$, $L = 5$, $\text{PNR} = 5 \;\text{dB}$, $N_t^{\text{beam}} = 24$ and $N_r^{\text{beam}} = 36$. It is interesting that the SE of any precoder decreases with an increasing number of common supports. This mainly comes from that the spatial degrees of freedom (DoFs)  are reduced due to less independent scatterers of the DPA-MIMO channels. Additionally, the SE curves of the proposed JOMP and JSBL-$\ell_2$ estimators approach that of the DFTB estimator as the number of the common supports increases.

\begin{figure}[!htb]
\centering
\includegraphics[width=3.3in]{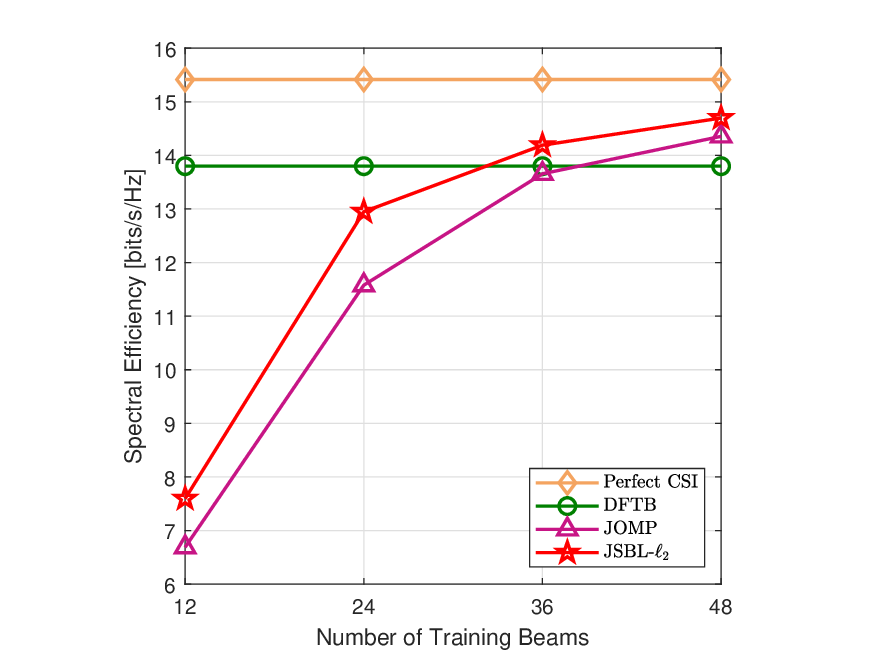}
\caption{SE versus number of training beams $N^{\text{beam}} = N_t^{\text{beam}} = N_r^{\text{beam}}$ when $M_t = M_r = 4$, $N_t^{\text{sub}} = N_r^{\text{sub}} = 12$, $N_s = 4$, $L = 5$, $L_c = 3$, $\text{DNR} = 5 \;\text{dB}$ and $\text{PNR} = 5 \;\text{dB}$.}
\label{fig_rate_csi_vs_beam}
\end{figure}
In Fig.{~\ref{fig_rate_csi_vs_beam}}, we depict the SE versus the number of training beams for a DPA-MIMO setup with $M_t = M_r = 4$, $N_t^{\text{sub}} = N_r^{\text{sub}} = 12$, $N_s = 4$, $L = 5$, $L_c = 3$, $\text{DNR} = 5 \;\text{dB}$ and $\text{PNR} = 5 \;\text{dB}$. The SE of the proposed JOMP and JSBL-$\ell_2$ estimators evidently boost towards the perfect CSI case with an increasing number of training beams. Furthermore, the SE of the proposed two estimators approximates that of the DFTB estimator (full-training case) with only its $\frac{3}{4} \times \frac{3}{4} = 56.25\% $ training overhead.

\section{Conclusion}
In this paper, we have focused on modeling and analysis of the narrowband DPA-MIMO based transceiver system, and designed efficient channel estimation and hybrid precoding schemes for such distributed array-of-sub-arrays architecture working on mmWave bands. Based on the reasonable analysis in Section II, the DPA-MIMO channel has high probability to manifest a hidden structured sparsity in the beam-domain channel vector due to the partially shared scatterers among the distributed sub-arrays at mmWave frequencies at the TX/RX. In light of this characteristic, we have formulated a structured SMV problem that estimates the AoDs, AoAs and the corresponding gain of significant paths. In order to guarantee the good recovery performance and decrease the training feedback overhead, the open-loop training beam patterns were designed through minimizing the total coherence of the equivalent measurement matrix. The simulation and comparison results have demonstrated that the proposed channel estimators can better exploit the structured channel properties defined in Definition~{\ref{def_dpa_mimo_chl}} than the existing CS-based estimators such as the OMP and SBL estimators, and the proposed hybrid precoding method enjoys the low-complexity while achieving good performance. 

More realistic channel modeling for DPA-MIMO is expected for future research. Specifically, the channel parameters including common and local scattering components can be investigated by using ray-tracing tools, and the scatter evolution on both sub-array and time axes should be also taken into account~\cite{wu2015non}. In addition, interesting and practical topics in DPA-MIMO applications cover many diversified situations, such as multi-user channel acquisition and hybrid precoding schemes~\cite{raghavan2017single}, and optimal sub-array placement for preventing blockage~\cite{raghavan2019antenna}.

\begin{appendices}
\section{Derivation of (\ref{equ_ySigmay})}
\label{appendix_derivation}
Let $\mathbf{s} = \mathbf{x} - \mathbf{c}$, $\mathbf{A} = \left( \boldsymbol{\Gamma }^{s} \right )^{-1}$ and $\mathbf{B} = \left( \mathbf{I}_{K} \otimes \boldsymbol{\Gamma }^{c} \right )^{-1}$, thus we can transform the objective function in (\ref{equ_ySigmay}) to the following 
\begin{equation}
\label{equ_fun_xc}
\underset{\mathbf{x},\mathbf{c}}{\mathrm{min}}\frac{1}{\lambda} \left \| \mathbf{y} - \boldsymbol{\Phi} \mathbf{x} \right \|_{2}^{2} + \left (\mathbf{x} - \mathbf{c}  \right )^{H} \mathbf{A} \left (\mathbf{x} - \mathbf{c}  \right ) + \mathbf{c}^{H} \mathbf{B} \mathbf{c}.  
\end{equation}
For the fixed $\mathbf{x}$, we have an unconstrained quadratic function only with respect to $\mathbf{c}$ and get its optimal solution as $\mathbf{c}^{\star } = \left ( \mathbf{A} + \mathbf{B}\right )^{-1} \mathbf{A}  \mathbf{x}$. 
After submitting the optimal $\mathbf{c}^{\star }$ into (\ref{equ_fun_xc}), we have 
\begin{equation}
\label{equ_fun_x}
\begin{aligned}
& \underset{\mathbf{x}}{\mathrm{min}}\frac{1}{\lambda} \left \| \mathbf{y} - \boldsymbol{\Phi} \mathbf{x} \right \|_{2}^{2} + \mathbf{x}^{H } \left ( \mathbf{A} - \mathbf{A}\left ( \mathbf{A} + \mathbf{B} \right )^{-1}\mathbf{A} \right )\mathbf{x}  \\
& \overset{\left( a \right) }{=}  \underset{\mathbf{x}}{\mathrm{min}}\frac{1}{\lambda} \left \| \mathbf{y} - \boldsymbol{\Phi} \mathbf{x} \right \|_{2}^{2} + \mathbf{x}^{H } \left ( \mathbf{A}^{-1} + \mathbf{B}^{-1} \right )^{-1}\mathbf{x} \\
& \overset{\left( b \right) }{=}  \mathbf{y}^{H } \left ( \lambda \mathbf{I}_{N}+ \boldsymbol{\Phi} \left ( \mathbf{A}^{-1} + \mathbf{B}^{-1} \right )\boldsymbol{\Phi}^{H} \right )^{-1}\mathbf{y},
\end{aligned}
\end{equation}
where (a) follows from the Woodbury identity~\cite{kaare2000cookbook}, and (b) follows from the identity \cite{wipf2011latent}
\begin{equation}
\mathbf{y}^{H } \left ( \lambda \mathbf{I}_{N}+ \boldsymbol{\Phi}\boldsymbol{\Gamma }\boldsymbol{\Phi}^{H} \right )^{-1}\mathbf{y} = \underset{\mathbf{x}}{\mathrm{min}} \; \frac{1}{\lambda} \left \| \mathbf{y} - \boldsymbol{\Phi} \mathbf{x} \right \|_{2}^{2} + \mathbf{x}^{H} \boldsymbol{\Gamma }^{-1} \mathbf{x},
\end{equation}
with $\boldsymbol{\Gamma} = \mathbf{A}^{-1} + \mathbf{B}^{-1} $. 

Obviously, the optimal value of $\mathbf{x}$ is expressed as
\begin{equation}
\begin{aligned}
\mathbf{x}^{\star} & = \left (  \lambda\boldsymbol{\Gamma }^{-1}+ \boldsymbol{\Phi}^H\boldsymbol{\Phi}\right )^{-1} \boldsymbol{\Phi}^H\mathbf{y}\\
& = \boldsymbol{\Gamma} \boldsymbol{\Phi}^H \left ( \lambda \mathbf{I}_{N}+ \boldsymbol{\Phi}\boldsymbol{\Gamma }\boldsymbol{\Phi}^{H} \right )^{-1}\mathbf{y}.
\end{aligned}
\end{equation}
Given the following matrix identity~\cite{kaare2000cookbook}
\begin{equation}
\left ( \mathbf{A} + \mathbf{B}\right )^{-1} \mathbf{A} \left (  \mathbf{A}^{-1} + \mathbf{B}^{-1} \right ) = \mathbf{B}^{-1},
\end{equation}
we further obtain 
\begin{subequations}
\begin{align}
\mathbf{s}^{\star } &= \mathbf{A}^{-1} \boldsymbol{\Phi}^H \left ( \lambda \mathbf{I}_{N}+ \boldsymbol{\Phi}\boldsymbol{\Gamma }\boldsymbol{\Phi}^{H} \right )^{-1}\mathbf{y},\\
 \mathbf{c}^{\star } & = \mathbf{B}^{-1} \boldsymbol{\Phi}^H \left ( \lambda \mathbf{I}_{N}+ \boldsymbol{\Phi}\boldsymbol{\Gamma }\boldsymbol{\Phi}^{H} \right )^{-1}\mathbf{y}.
\end{align}
\end{subequations}

\section{Iterative computation of $\mathbf{T}_{n}$ in \eqref{problem_opt_analog_structure}}
\label{appendix_derivation_2}
In order to  compute $\mathbf{T}_{n}$ iteratively, we further express $\mathbf{R}_n$ as
$\mathbf{R}_n = \mathbf{R}_{n-1} + \frac{P_{\mathrm d}}{N_s \sigma _{\mathrm z}^{2}}  \mathbf{H}  \tilde{\mathbf{v}}_{n} \tilde{\mathbf{v}}_{n}^{H} \mathbf{H}^{H}.$
With the Sherman-Morrison formula~\cite{kaare2000cookbook}
\begin{equation}
\left (\mathbf{A}+\tau \mathbf{x}\mathbf{x}^{H}  \right )^{-1} = \mathbf{A}^{-1} - \frac{\mathbf{A}^{-1} \tau \mathbf{x}\mathbf{x}^{H} \mathbf{A}^{-1}}{1 +\tau \mathbf{x}^{H}\mathbf{A}^{-1} \mathbf{x} },
\end{equation}
we have
\begin{equation}
\label{equ_Rn_inv}
\mathbf{R}_n^{-1} = \mathbf{R}_{n-1}^{-1}- \frac{P_{\mathrm d}}{N_s \sigma _{\mathrm z}^{2} \left ( 1 + \rho _{n} \right )}  \mathbf{R}_{n-1}^{-1} \mathbf{H}  \tilde{\mathbf{v}}_{n} \tilde{\mathbf{v}}_{n}^{H} \mathbf{H}^{H} \mathbf{R}_{n-1}^{-1},
\end{equation}
where $\rho _{n} = \frac{P_{\mathrm d}}{N_s \sigma _{\mathrm z}^{2}} \tilde{\mathbf{v}}_{n}^{H} \mathbf{T}_{n-1} \tilde{\mathbf{v}}_{n}$. Thus, the matrix $\mathbf{T}_n$ can be iteratively obtained by
\begin{equation}
\label{equ_Tn}
\mathbf{T}_n = \mathbf{T}_{n-1} - \frac{P_{\mathrm d}}{N_s \sigma _{\mathrm z}^{2} \left ( 1 + \rho _{n} \right )}   \mathbf{T}_{n-1} \tilde{\mathbf{v}}_{n}  \tilde{\mathbf{v}}_{n} ^{H} \mathbf{T}_{n-1}.
\end{equation}
\end{appendices}

\section*{Acknowledgment}
The authors would like to thank the Associate Editor and the anonymous reviewers for their valuable comments and helpful suggestions.

{
\bibliographystyle{IEEEtran}
\bibliography{IEEEabrv,mybibfile}
}
\ifCLASSOPTIONcaptionsoff
  \newpage
\fi

\end{document}